\documentclass[a4paper,11pt]{article}
\usepackage{jheppub} 
\usepackage{lineno}
\usepackage{hyperref}
\usepackage{latexsym,bm,amsmath,amssymb}
\usepackage{bbm}
\usepackage{physics}
\usepackage{color}
\usepackage{tikz}
\usepackage{tikz-feynman}
\usetikzlibrary{arrows}
\usepackage{graphicx}
\usepackage{multirow}
\usepackage{mathtools}
\usepackage{cancel}
\usepackage{diagbox}
\usepackage{makecell}
\usetikzlibrary{arrows.meta, calc}
\usepackage{subcaption}
\usepackage{caption}
\usepackage{amsmath}

\title{\boldmath Towards Bulk Locality: A Systematic Construction of Contact Interactions from Chord Diagrams}

\author[a]{Hao Dai$^{*}$,}
\affiliation[a]{Department of Physics, Shanghai University, Shanghai, 200444, China}
\author[b]{Yi-Li Wang$^{*}$,}
\affiliation[b]{Asia Pacific Center for Theoretical Physics, 77 Cheongam-ro, Pohang, 37673, Korea}
\author[c]{Yu-Ge Chen$^{*}$,}
\affiliation[c]{Institute for Quantum Science and Technology, Shanghai University, Shanghai 200444, China}
\author[d]{Li-Guo Qin,}
\affiliation[d]{School of Mathematics, Physics and Statistics, Shanghai University of Engineering Science, Shanghai, 201620, China}
\author[a]{Li-Jun Tian}
\emailAdd{isaac\_h\_dai@shu.edu.cn}
\emailAdd{yili.wang@apctp.org}
\emailAdd{chenyuge@shu.edu.cn}
\emailAdd{lgqin@foxmail.com}
\emailAdd{tianlijun@shu.edu.cn}

\abstract{
Chord diagrams encode boundary correlators in the double-scaled holographic Sachdev–Ye–Kitaev model, but currently capture only a limited class of bulk interactions that yield pure power-law correlators.
In this article, we investigate a general construction based on Fock-space flux models with arbitrary periodic lattice size, clarifies how lattice dimensions control probe configurations and bulk contact vertices. Developing a systematic matching scheme and using the chord path integral formalism, we compute three- to six-point contact correlators and reproduce a broad class of AdS$_2$ scalar contact Witten diagrams, including those with logarithmic singularities.
The results demonstrate that chord diagrams, in full generality, provide a microscopic description of bulk contact interactions and thereby establish a principled framework for reconstructing bulk locality from boundary data.
}

\begingroup

\footnotetext{$^{*}$These authors contributed equally to this work.}
\endgroup

\begin{document}
\maketitle
\flushbottom

\section{Introduction}
\label{sec:introduction}

\par Being a solvable strongly correlated system at large-$N$, the Sachdev-Ye-Kitaev (SYK) model \cite{Sachdev:1992fk,Kitaev2015,Maldacena:2016hyu,Chowdhury:2021qpy} has sparked
significant scientific interest. Its variants, such as Yukawa-SYK theory, offer a powerful aid in exploring the nature of strange metals \cite{Esterlis2021,Guo2022,Patel2023,Wang:2024utm,Sin:2025sue}, which is an important subfield in condensed mattter theories. Meanwhile, SYK-inspired theories also provide a promising avenue of exploring the microscopic mechanisms of quantum holography and the emergence of spacetime geometry from quantum many-body systems \cite{Maldacena:2016upp,Polchinski:2016xgd,Gu:2019jub,Garcia-Garcia:2020cdo}. A typical example is the double-scaled SYK (DSSYK) \cite{Cotler:2016fpe,Berkooz:2018jqr,Lin:2022rbf}, where the number of fermion interactions $p$ scales as $\sqrt{N}$. It is a theory for $1$D nearly conformal field theory (NCFT$_1$), and the model becomes analytically tractable through the combinatorics of chord diagrams \cite{Parisi1994,Flajolet2000,Berkooz2025}, which represent sequences of Hamiltonian insertions. The chords are conjectured to correspond to spacetime processes in a dual near-AdS$_2$ (NAdS$_2$) geometry \cite{Berkooz:2020uly,Berkooz:2018jqr,Lin:2022rbf}. Therefore, DSSYK thoery offers a concrete, albeit discrete, microscopic picture of holographic duality in the simplest setting of AdS$_2$/CFT$_1$ (NAdS$_2$/CFT$_1$).

\par Recent development shows that Parisi's hypercube \cite{Parisi1994}, without $p$-locality, shares similar combinatorial solution (and thus dynamics) of a DSSYK model \cite{Jia2020}. Such a similarity, along with 
chord path integral \cite{Jia2020,Berkooz:2024evs,Berkooz:2024ofm}, has enabled the computation of correlation functions beyond the original DSSYK model. 
Certain contact chord diagrams, arising from a broader class of Fock-space flux models, can reproduce the correlators of some scalar contact Witten diagrams in AdS$_2$ \cite{Berkooz:2023scv,Berkooz:2023cqc,Jia:2024tii,Jia:2025tvn}. Therefore, interpretating of chords as fundamental building blocks of bulk locality is well motivated. However, current approaches remain limited in several important respects. First, they focus mainly on a single, alternating sign configuration of probe operators, which corresponds to a specific type of bulk contact vertex. Second, the correlators obtained are purely power-law functions of the boundary coordinates, whereas generic AdS$_2$ contact Witten diagrams often contain logarithmic singularities \cite{Bliard:2022xsm} -- a hallmark of more intricate bulk couplings or derivative interactions. Additionally, no systematic framework exists for combining chord diagram configurations to describe general bulk contact interactions.

While Ref.~\cite{Jia:2025tvn} established the basic three- and four-point contact diagram construction on fixed lattices, the present work generalizes the framework to arbitrary $L$, incorporates the linear combination and parameter degeneracy mechanisms, and extends the matching to five- and six-point correlators including logarithmic and rational cases.
\par In this paper, we address these limitations and make substantial progress towards reconstructing bulk locality from chord diagrams. We present a generalized construction of contact chord diagrams based on Fock-space flux models with arbitrary lattice sizes $L$. The lattice size $L$ controls the set of allowed probe sign configurations (or ``paths'') and thus the variety of contact vertices that can be represented. We suggest that the physical correlation functions have to be a \emph{linear combination} of contributions from all such allowed configurations. To generate the logarithmic singularities in Witten diagrams, we apply the \emph{parameter degeneracy} technique. By tuning the flux parameters to render diagrammatic exponents nearly degenerate, and taking a controlled limit with suitably diverging weights, one obtains terms of the form $x^a \log x$ or $(1-x)^b \log(1-x)$.

\par This generalised framework, together with the chord path integral, enables the computation of three- to six-point contact correlators. We systematically match these results to the known library of AdS$_2$ scalar contact Witten diagrams \cite{Bliard:2022xsm}, including cases with logarithmic singularities that were previously inaccessible. These results indicates that chord diagrams, in full generality, suffice to describe a wide class of bulk contact interactions, suggesting that chord diagrams may provide the basis for a more systematic framework for reconstructing bulk locality from boundary data.

\par This article is organized as follows. In Section~\ref{sec:model}, we review the Fock-space flux models and generalize the construction to arbitrary periodic lattice sizes $L$. Section~\ref{sec:CFT} briefly recalls the chord path integral formalism, which is the main computational tool used throughout this work. Section~\ref{sec:correlators} contains our core calculations: we compute three- to six-point contact correlators and match them with AdS$_2$ scalar contact Witten diagrams. In Section~\ref{sec:linear_comb}, we present the linear combination method and the parameter degeneracy mechanism that generates logarithmic singularities. We conclude with implications for holography and bulk locality, and outline future directions in Section~\ref{sec:discussion}.

\section{Fock-Space Flux Models and General Periodic Lattices}
\label{sec:model}

In this section, we review the construction of Fock-space flux models that generalize the double-scaled SYK model, and extend it to accommodate arbitrary periodic lattice sizes $L$. These models provide the microscopic foundation for the contact chord diagrams we will study.

\subsection{A minimal introduction to the model}
\par We will start with a brief introduction to the Fock-space flux models based on refs. \cite{Berkooz:2023cqc,Berkooz:2023scv,Jia:2025tvn}. For more details on chord diagrams, we refer to the nice review by Berkooz and Mamroud \cite{Berkooz2025}.\\

The starting point is a set of $N$ ``Fock-space oscillators'' $\mathbb{S}_\nu^\pm$ for each direction $\mu = 1,\dots,N$ and a center element operator $\mathbb{S}_\mu^3$, which can be thought of as generators of a magnetic algebra \cite{Berkooz:2023scv}. The oscillators satisfy the commutator
\begin{equation}
	[\mathbb{S}_\mu^3, \mathbb{S}_\nu^\pm] = \pm \delta_{\mu\nu} \mathbb{S}_\mu^\pm.
\end{equation}
In Parisi's original setup, there are superconducting dots on the hypercube vertices, whilst the energy is frustated by a disordered magnetic flux. Following Refs.\cite{Berkooz:2023cqc,Berkooz:2023scv,Jia:2025tvn}, we do not consider superconductivity, but modify the physical picture into a particle hopping on these $2^N$ vertices, and the magnetic flux will affect the return amplitudes of the hopping particle.
The Hamiltonian takes the form \cite{Jia:2025tvn}
\begin{equation}\label{eqn:hamiltonian}
H = \frac{c}{\sqrt{N}} \sum_{\mu=1}^N D_\mu, \qquad D_\mu = T_\mu^+ + T_\mu^-,
\end{equation}
where the constant $c$ is chosen such that $\langle \operatorname{Tr} H^2 \rangle = L^N = \text{Hilbert space dimension}$, where $L$ is the dimension of the representation of $\mathbb{S}^\pm$ and $\mathbb{S}^3$, and $L$ gives the number of sites along each direction of a hypercubic lattice in Fock space. \\
In a similar vein to ref.\cite{Jia:2025tvn}, we define the fluxed operators
\begin{equation}
T_\mu^\pm = \mathbb{S}_\mu^\pm \prod_{\rho \neq \mu} e^{\pm \frac{i}{2} F_{\mu\rho} \mathbb{S}_\rho^3},
\end{equation}
where $F_{\mu\nu}$ denotes random fluxes (antisymmetric, independently and identically distributed for each unordered pair $\{\mu,\nu\}$). Moreover,
we require $\langle\sin F\rangle=0$ and $q:=\langle\cos F\rangle$, where $\langle...\rangle$ denotes disorder averaging, and $q$ is a tunable parameter.

These satisfy the magnetic algebra
\begin{equation}
	T_\mu^+ T_\nu^+ = e^{i F_{\mu\nu}} T_\nu^+ T_\mu^+, \qquad
	T_\mu^+ T_\nu^- = e^{-i F_{\mu\nu}} T_\nu^- T_\mu^+.
	\label{eq:magnetic algebra}
\end{equation}
 we consider finite-dimensional representations of these operators. This Hamiltonian describes the hopping of a particle on the vertices, and $T^{\pm}_\mu$ is a transport operator which enables the particle to move long the positive/negative $\mu$ direction (with a phase term). \footnote{In this model, the hopping happens between spin configurations, not between the real-space postion.}\\

The next step is to define probe operators that are suitable observalbes in this model. We assume that our theory \eqref{eqn:hamiltonian} is an IR correspondence to a UV theory via AdS/CFT correspondence. When flowing to IR, observables in UV will become very complicated (highly scrambled). Since our IR theory is chaotic, generic complicated operators statistically resemble the Hamiltonian. Therefore, a natural phenomenological choice for probes is an ensemble of operators with the same structural and statistical properties as eqn.\eqref{eqn:hamiltonian} \cite{Berkooz:2018qkz,Berkooz:2023cqc}. In other words, the probes should be ``single-trace'' operators.\\

To this end, we introduce another set of fluxes $\tilde{F}_{\mu\nu}$ and define $\tilde{T}^\pm$ similarly \cite{Jia:2025tvn},
\begin{eqnarray}
    \tilde{T}_\mu^{(i)\pm} = \mathbb{S}_\mu^\pm \prod_{\rho \neq \mu} e^{\pm \frac{i}{2} \tilde{F}_{\mu\rho}^{(i)} \mathbb{S}_\rho^3}.
\end{eqnarray}
A simple choice over probe operators $O_i$ ($i=1,\dots,n$) can be defined with independent fluxes $\tilde{F}_{\mu\nu}^{(i)}$ (which may be correlated with $F_{\mu\nu}$) \cite{Jia:2025tvn}, similar to eqn.\eqref{eqn:hamiltonian},
\begin{equation}
O_i = \frac{c_i}{\sqrt{N}} \sum_{\mu=1}^N \tilde{D}_\mu^{(i)}, \qquad \tilde{D}_\mu^{(i)} = \tilde{T}_\mu^{(i)+} + \tilde{T}_\mu^{(i)-}.
\end{equation}
The constants $c_i$ will normalize the two-point functions appropriately. The fluxes satisfy
\begin{equation}
	T_{\mu}^{ \pm} \tilde{T}_{\nu}^{+}=e^{ \pm i \frac{F_{\mu \nu}+\tilde{F}_{\mu \nu}}{2}} T_{\nu}^{+} T_{\mu}^{+}, \quad T_{\mu}^{ \pm} \tilde{T}_{\nu}^{-}=e^{\mp i \frac{F_{\mu \nu}+\tilde{F}_{\mu \nu}}{2}} \tilde{T}_{\nu}^{-} T_{\mu}^{+}.
	\label{eq:magnetic algebra2}
\end{equation}

To maximize the range of phenomena that the model can address, we introduce periodic boundary conditions and define cyclic raising/lowering operators
\begin{equation}
	P_\mu^+ = \mathbb{S}_\mu^+ + (\mathbb{S}_\mu^-)^{L-1}, \qquad P_\mu^- = (P_\mu^+)^\dagger,
\end{equation}
{which satisfy $(P_{\mu}^{+})^{L} = 1$. 
The representation of $\mathbb{S}_{\mu}^{3}$ is chosen to be diagonal, with eigenvalues
\begin{equation}
    s_m = \frac{L+1}{2} - m, \qquad m = 1, 2, \dots, L,
    \label{eq:S3_eigenvalues}
\end{equation}
so that $\mathbb{S}_{\mu}^{3} = \operatorname{diag}\bigl(\frac{L-1}{2}, \frac{L-3}{2}, \dots, -\frac{L-1}{2}\bigr)$. 
For odd $L$, these eigenvalues are integers; for even $L$, they are half-integers.
The operators $\mathbb{S}_{\mu}^{\pm}$ are defined by $(\mathbb{S}^{+})_{m,m+1}=1$ for $m=1,\dots,L-1$ and $\mathbb{S}^{-}=(\mathbb{S}^{+})^{\dagger}$. } To be specific, $\mathbb{S}_\mu^\pm$ take the form
\[
\mathbb{S}^+ = \begin{pmatrix}
	0 & 1 & 0 & \cdots & 0 \\
	0 & 0 & 1 & \cdots & 0 \\
	\vdots & \vdots & \ddots & \ddots & \vdots \\
	0 & 0 & \cdots & 0 & 1 \\
	0 & 0 & \cdots & 0 & 0
\end{pmatrix}, \quad
\mathbb{S}^- = \begin{pmatrix}
	0 & 0 & 0 & \cdots & 0 \\
	1 & 0 & 0 & \cdots & 0 \\
	0 & 1 & 0 & \cdots & 0 \\
	\vdots & \ddots & \ddots & \ddots & \vdots \\
	0 & \cdots & 0 & 1 & 0
\end{pmatrix}.
\]
To guarantee the relations \eqref{eq:magnetic algebra}, the fluxes should satisfy the constraint
\begin{equation}
	e^{i \frac{F}{2} \mathbb{S}^{3}} P^{ \pm} e^{-i \frac{F}{2} \mathbb{S}^{3}}=e^{ \pm i \frac{F}{2}} P^{ \pm}.
\end{equation}
It turns out that the fluxes can only take the form
\begin{equation}\label{eq:flux_quantization}
	F=\frac{4 n \pi}{L}, \quad n \in \mathbb{Z}.
\end{equation}
The same discussion applies to $\tilde{F}$. For $L=2$ there is no constraint on the fluxes, while for $L\geqslant3$ the allowed values of the fluxes increase as $L$ grows. \\

\par {We are interested in partition functions and $n$-point correlation functions, which require handling expressions of the form $\langle \operatorname{Tr} H^{k} \rangle$ and
$\langle \operatorname{Tr}( H^{k_n} \tilde{D}^{(n)}_\mu H^{k_{n-1}} \dots \tilde{D}^{(1)}_\mu ) \rangle$. In the large-$N$ limit, the leading-order contribution to the moments of $H$ 
comes from pairwise contractions of the Fock-space indices $\mu=1,\dots,N$, where each index appears exactly twice. 
Explicitly, the leading moments are given by sums such as
\begin{equation}
    \langle \operatorname{Tr} H^4 \rangle_{\text{leading}} = 
    \frac{\text{constant}^2}{N^2}\sum_{\mu\neq\nu} 
    \langle \operatorname{Tr}(D_\mu D_\mu D_\nu D_\nu 
    + D_\mu D_\nu D_\mu D_\nu 
    + D_\mu D_\nu D_\nu D_\mu) \rangle,
\end{equation}
which reproduce the DSSYK moments after averaging over the random fluxes.}

{Subleading contributions arise when an index appears more than twice. 
For instance, a cubic moment of the Hamiltonian is of the form
\begin{equation}
    \langle \operatorname{Tr} H^3 \rangle = 
    \frac{\text{constant}^3}{N^{3/2}}\sum_{\mu} 
    \langle \operatorname{Tr}(D_\mu D_\mu D_\mu) \rangle.
\end{equation}
Both the leading and subleading structures of the Fock-space flux model were established in Refs.~\cite{Berkooz:2023cqc,Berkooz:2023scv}. 
Recent research~\cite{Jia:2025tvn} suggests that subleading interactions among probe operators may correspond to contact interactions. }
Therefore, we consider $n$-point contact contributions 
where all probe operators act on the same lattice index $\mu$,
\begin{equation}
	\left\langle O_{n}\left(\tau_{n}\right) O_{n-1}\left(\tau_{n-1}\right) \cdots O_{1}\left(\tau_{1}\right)\right\rangle,
\end{equation}
The relevant moments for Euclidean correlators are of the form \cite{Jia:2025tvn}
\begin{equation}
M_{k_1,\dots,k_n}^{\text{contact}} = \frac{1}{N^{(n + \sum_i k_i)/2}} \sum_\mu 
\left\langle \operatorname{Tr} \left( H^{k_n} \tilde{D}_\mu^{(n)} H^{k_{n-1}} \cdots H^{k_1} \tilde{D}_\mu^{(1)} \right) \right\rangle,
\end{equation}
where \(\langle ... \rangle\) denotes, averaging over the fluxes $F$ and $\tilde{F}^{(i)}$.

{In the geometric picture established in Refs.~\cite{Berkooz:2023cqc,Berkooz:2023scv,Jia:2025tvn},
expanding each $H$ and $\tilde{D}_\mu^{(i)}$ into sums of $T^\pm$ and $\tilde{T}^{(i)\pm}$ operators yields monomials that correspond to sequences of hopping steps on the $N$-dimensional hypercubic lattice with $L$ sites per direction: $T_\mu^+$ and $\tilde{T}_\mu^{(i)+}$ move the particle one step forward along the $\mu$-direction, while $T_\mu^-$ and $\tilde{T}_\mu^{(i)-}$ move it one step backward, each step accumulating a phase factor from the fluxes in the transverse directions. The trace imposes that the overall path must close;  we therefore analyze the relation between closed paths and the dimension $L$ in the next subsection.}

\subsection{ closure condition and allowed configurations}
\label{sec:closure}
For contact contributions, all probe operators act on the same $\mu$, so the path is non-trivial only in the $\mu$-direction, while in other directions only phase factors from the exponentials contribute. Closure in the $\mu$-direction requires that the net displacement (sum of $\pm 1$ steps from $P_\mu^\pm$) is zero modulo $L$. This imposes a condition on the choice of signs $\eta_i = \pm 1$ for each $\tilde{T}_\mu^{(i)\eta_i}$ and also on the signs from any $T_\mu^\pm$ that might appear from $H$ expansions. {However, in the chord diagram representation, the $n$ probe operators are drawn as chords that meet at a common point in the interior of the disk---the contact point---while $H$-chords are drawn as arcs that do not attach to this contact point~\cite{Jia:2025tvn}. The $H$-chords represent $T^\pm$ operators acting on directions other than $\mu$. }Therefore, for a pure contact contribution without $H$ operators acting on the $\mu$ direction, we require:
\begin{equation}
    \sum_{i=1}^n \eta_i \epsilon_i = 0 \mod L,
    \label{eq:closure_condition}
\end{equation}
where $\epsilon_i$ is the intrinsic sign of the probe operator (the superscript of $\tilde{T}^{(i)\pm}$) and $\eta_i$ is the sign chosen in the expansion. 
Equivalent constraints have been recognized, though not always stated in this explicit mathematical form, in the Fock-space flux construction of Refs.~\cite{Berkooz:2023cqc,Berkooz:2023scv} and in the contact diagram analysis of Ref.~\cite{Jia:2025tvn}. 
Here we formulate the condition precisely and systematically use it to determine the allowed sign configurations for arbitrary $n$ and $L$, which is the foundation of the generalized matching scheme developed in this work.

\par The lattice size $L$ plays a crucial role in determining the set of allowed probe sign configurations. 
For a given $n$, we must enumerate all sign sequences $\{\epsilon_i = \pm 1\}$ that satisfy 
$\sum_i \epsilon_i = 0 \mod L$. 
{This condition is necessary because the trace forces the overall path on the Fock-space lattice to close; 
any sequence violating it would correspond to an open path and thus give a vanishing contribution. In the pure contact limit---where no $H$-chord acts on the $\mu$-direction---\eqref{eq:closure_condition} 
with $\eta_i=\epsilon_i$ is also sufficient: once the net displacement vanishes modulo $L$, the trace evaluates to a non-zero factor, as we will compute explicitly in Section~\ref{sec:correlators}.} The number of such sequences depends on $L$. We denote the set of configurations by $\mathcal{C}$. Take four points ($n=4$) for example:
\begin{itemize}
\item {For $L=2$, the condition $\sum_i \epsilon_i = 0 \mod 2$ requires an even number of $+1$'s (and hence an even number of $-1$'s). For $n=4$, this allows all configurations with zero, two, or four $+1$'s: 
$(+,-,+,-)$, $(-,+,-,+)$, $(+,+,-,-)$, $(-,-,+,+)$, $(+,-,-,+)$, $(-,+,+,-)$, as well as $(+,+,+,+)$ and $(-,-,-,-)$.}
\item For $L=3$, the condition allows sequences where the sum is a multiple of 3. For $n=4$, this includes all six permutations of two $+$ and two $-$ signs, because $2-2=0$ which is divisible by 3. It also allows sequences like $(+,+,+,+)$? No, because $4$ mod $3 = 1 \neq 0$. So only two positive and two negative signs work for $n=4$.
\item For $L=4$, sequences with sum $0,\pm4,\pm8,\dots$ mod 4 are allowed. For $n=4$, this includes all two-positive-two-negative permutations (sum 0), as well as all-positive $(+,+,+,+)$ (sum 4) and all-negative $(-,-,-,-)$ (sum -4) because $4$ and $-4$ are divisible by 4.
\end{itemize}
Table~\ref{tab:configs_L} summarizes the allowed four-point configurations for small values of $L$.
\begin{table}[htbp]
	\centering
	\caption{Allowed four-point probe sign configurations for different lattice sizes $L$. The configurations are listed up to overall sign reversal (all $+ \leftrightarrow -$).}
	\label{tab:configs_L}
	\begin{tabular}{c|c|c}
		\hline
		Lattice size $L$ & Allowed configurations for $n=4$ & Number of configurations  \\
		\hline
		2 & $(+,+,+,+)$, $(+,+,-,+)$, $(+,-,+,-)$, $(+,-,-,+)$ & 4 \\
		3 & $(+,+,-,-)$, $(+,-,+,-)$, $(+,-,-,+)$ & 3 \\
		4 & $(+,+,+,+)$, $(+,+,-,-)$, $(+,-,+,-)$, $(+,-,-,+)$ & 4 \\
		\hline
	\end{tabular}
\end{table}

Thus, by varying $L$, we can access different sets of probe sign configurations, each of which may correspond to a distinct type of bulk contact vertex.  Next, we consider the effect of a single configuration on the fluxes. Inserting a specific configuration into the trace, i.e., computing an expression such as$\operatorname{Tr}\left(\cdots \tilde{T}_{\mu}^{-} \cdots \tilde{T}_{\mu}^{-} \cdots \tilde{T}_{\mu}^{+} \cdots \tilde{T}_{\mu}^{-}\right)$, yields a factor
\begin{equation}\label{eq:trace_factor}	 
\Bigl\langle \operatorname{Tr}\!\Bigl[ \exp\!\Bigl( \frac{i}{2}\sum_{i=1}^n \epsilon_i \tilde F^{(i)} \mathbb{S}^3 \Bigr) \Bigr] \Bigr\rangle^{N-1},
\end{equation}

In the large-$N$ limit, in order for the contact contribution not to be exponentially suppressed after averaging over fluxes, the fluxes must satisfy a linear constraint. This constraint arises from summing over the phases from the directions $\rho \neq \mu$. Generally, it takes the form
\begin{equation}\label{eq:universal_constraint}
\sum_{i=1}^n c_i \tilde{F}^{(i)} \in 4\pi \mathbb{Z},
\end{equation}
For the alternating configuration $(+,-,+,-)$, $c_i = \epsilon_i$ \cite{Berkooz:2023cqc,Berkooz:2023scv}, i.e.
$\tilde{F}^{(1)} - \tilde{F}^{(2)} + \tilde{F}^{(3)} - \tilde{F}^{(4)} \in 4\pi \mathbb{Z}$.

As established in this section, the lattice size $L$ in the Fock-space representation determines the closure condition for probe operators: $\sum_i \epsilon_i = 0 \mod L$. For a fixed number of points $n$, different values of $L$ allow for different sets of sign sequences $\{\epsilon_i\}$. Each allowed sequence can be thought of as a distinct ``path'' in the space of probe insertions, and each such path corresponds to a particular pattern of chord intersections in the diagrammatic expansion. Crucially, we propose that different probe sign configurations correspond to different types of contact vertices in the bulk dual theory. Thus, by varying $L$, we can access a richer set of bulk contact interactions.

\par We have presented the Fock-space flux model generalizing the DSSYK model, with an emphasis on the role of the lattice size $L$. The condition $\sum_i \epsilon_i = 0 \mod L$ selects the allowed probe sign configurations, and additional flux constraints are necessary to avoid exponential suppression. In the following, we will use this framework to build a general theory of contact chord diagrams.

\section{Conformal form, Chord rules and Path integral}
\label{sec:CFT}
\par This section collects the necessary technical ingredients---conformal factors, chord rules, positivity constraints, and the chord path integral---following Refs.~\cite{Berkooz:2023cqc,Berkooz:2023scv,Berkooz:2024evs,Berkooz:2024ofm}. Readers familiar with these tools may skip to Section~\ref{sec:correlators}.

\par For each configuration $\mathcal{C} = \{\epsilon_i\}$, we can derive chord rules that determine the weight of a diagram in terms of the number of intersections between $H$-chords and probe chords. The basic principles are as in the original DSSYK model \cite{Berkooz:2018jqr}, but the factors associated with intersections depend on the signs of the probes involved.

When an $H$-chord (with intrinsic sign $\eta = \pm 1$) crosses a single probe chord of sign $\epsilon_i$, the phase accumulated is proportional to $\eta \epsilon_i (F + \tilde{F}^{(i)})/2$. After averaging over fluxes, this leads to a factor
\begin{equation}\label{eqn:single}
	q^{\Delta_i} = \left\langle \cos \frac{F + \tilde{F}^{(i)}}{2} \right\rangle.
\end{equation}
When an $H$-chord crosses two probe chords with signs $\epsilon_i$ and $\epsilon_j$, the accumulated phase is proportional to $\eta [\epsilon_i (F+\tilde{F}^{(i)})/2 + \epsilon_j (F+\tilde{F}^{(j)})/2]$. After averaging, we obtain a factor that depends on whether $\epsilon_i = \epsilon_j$:
\begin{equation}\label{eqn:two}
	q^{\Delta_{ij}} = \begin{cases}
		\displaystyle \left\langle \cos\left( F + \frac{\tilde{F}^{(i)} + \tilde{F}^{(j)}}{2} \right) \right\rangle, & \epsilon_i = \epsilon_j, \\[10pt]
		\displaystyle \left\langle \cos \frac{\tilde{F}^{(i)} - \tilde{F}^{(j)}}{2} \right\rangle, & \epsilon_i \neq \epsilon_j.
	\end{cases}
\end{equation}
In the conformal limit where the flux distributions are sharply peaked around zero, eqns.\eqref{eqn:single} and \eqref{eqn:two} become
\begin{equation}\label{eq:Delta_def}
\Delta_i \approx \frac{\langle (F+\tilde{F}^{(i)})^2 \rangle}{4\langle F^2 \rangle}, \qquad
	\Delta_{ij} \approx \begin{cases}
		\displaystyle \frac{\langle (F + (\tilde{F}^{(i)}+\tilde{F}^{(j)})/2)^2 \rangle}{4\langle F^2 \rangle}, & \epsilon_i = \epsilon_j, \\[10pt]
		\displaystyle \frac{\langle (\tilde{F}^{(i)} - \tilde{F}^{(j)})^2 \rangle}{4\langle F^2 \rangle}, & \epsilon_i \neq \epsilon_j.
	\end{cases}
\end{equation}

\textbf{Chord rules for general configurations.}

{Based on the above discussion, we can now formulate the chord diagram rules for a general $n$-point contact interaction.
Consider a fixed chord diagram with a given set of $H$-chords and probe chords meeting at the contact point.
The probe chords divide the boundary circle into $n$ intervals, and their cyclic order is fixed by the time ordering of the operators.
An $H$-chord connecting two boundary points can only traverse a set of probe intervals that are consecutive in this cyclic order; geometrically, it cannot skip a probe chord and then intersect a later one without also crossing the intermediate one.}

{With this adjacency condition, each $H$-chord contributes a weight determined exclusively by the contiguous set of probe operators whose intervals it crosses:
\begin{itemize}
    \item If an $H$-chord crosses the interval of a single probe operator $O_i$,
    it contributes a factor $q^{\Delta_i}$.
    \item If it crosses two adjacent probe operators $O_i$ and $O_{i+1}$ (with indices
    modulo $n$), it contributes a factor $q^{\Delta_{i,i+1}}$.
    \item If it crosses three adjacent probe operators $O_i, O_{i+1}, O_{i+2}$,
    it contributes a factor $q^{\Delta_{i,i+1,i+2}}$, and so on for larger contiguous
    blocks.
\end{itemize}
Each $H$-chord is assigned exactly one such factor---the one corresponding to the precise contiguous set of probe intervals it traverses. There is no product of single- and multi-probe factors for the same $H$-chord. $H$-$H$ intersections contribute the usual factor $q$, exactly as in the standard DSSYK model~\cite{Berkooz:2018jqr}.
The total weight of a diagram can therefore be written as
\begin{equation}
    q^{\# H\text{-}H \text{ int.}} \;
    \prod_{\text{contiguous } S\subseteq\{1,\dots,n\},\,S\neq\varnothing}
    \bigl(q^{\Delta_S}\bigr)^{\# H\text{-}(O_i,i\in S) \text{ int.}},
    \label{eq:general_chord_rule}
\end{equation}
where the product runs over all non-empty contiguous subsets of the cyclic index set
$\{1,\dots,n\}$, and $\Delta_S$ is the conformal factor associated with an $H$-chord
crossing all probe operators in $S$.} \\

\textbf{Parameter independence and flux constraints.}

{Equation~\eqref{eq:general_chord_rule} states the chord rules in their most general form. In practice, however, not all $\Delta_S$ are independent. The linear flux constraint $\sum_i\epsilon_i\tilde{F}^{(i)}\in 4\pi\mathbb{Z}$
implies relations among the multi-probe factors, whose precise form depends on $n$ and on the chosen sign configuration:
\begin{itemize}
    \item For $n=4$, one has $\Delta_{12}=\Delta_{34}$ and $\Delta_{23}=\Delta_{14}$.
    Thus among the four contiguous two-probe factors only two are independent
    (and for the alternating configuration this number reduces further).
    \item For $n=5$, contiguous three-probe factors like $\Delta_{123}$ are not
    independent parameters; e.g.\ for the configuration $(+,+,+,+,-)$ one finds
    $\Delta_{123}=\Delta_{45}$, so the three-probe factor is entirely determined
    by a two-probe one.
    \item For $n\ge 6$, genuine independent three-probe factors $\Delta_{ijk}$ appear
    (and, more generally, $k$-probe factors up to $k=\lfloor n/2\rfloor$ can become
    independent depending on the closure condition).
\end{itemize}
These relations guarantee that, despite the formal sum over contiguous subsets in \eqref{eq:general_chord_rule}, the number of independent flux parameters never exceeds the degrees of freedom of the underlying Gaussian flux distribution. In the explicit computations that follow, we always work with the minimal set of independent parameters appropriate to the specific $n$, $L$, and sign configuration.}

\textbf{Positivity constraints}

The parameters $\Delta_i$ and $\Delta_{ij}$ are not independent; they must arise from an underlying Gaussian distribution of the fluxes. 
Defining the centered variables
\begin{equation}
    a_i = \frac{F + \tilde{F}^{(i)}}{2\sqrt{\langle F^2 \rangle}}, \qquad i=1,\dots,n,
    \label{eq:a_i}
\end{equation}
we have $\Delta_i = \langle a_i^2 \rangle$. 
The two-index parameters are given by
\begin{equation}
    \Delta_{ij} = 
    \begin{cases}
        \langle (a_i + a_j)^2 \rangle, & \epsilon_i = \epsilon_j, \\[6pt]
        \langle (a_i - a_j)^2 \rangle, & \epsilon_i \neq \epsilon_j .
    \end{cases}
    \label{eq:Delta_ij_from_a}
\end{equation}
The necessary and sufficient condition for physical realizability 
is that the covariance matrix $M_{ij} = \langle a_i a_j \rangle$ be positive semi-definite.
This single requirement encodes all constraints on the $\Delta$ parameters.

{A set of necessary (but generally not sufficient) conditions follows from the positivity 
of the $2\times2$ principal minors of $M$, yielding the pairwise inequalities
\begin{equation}
    (\sqrt{\Delta_i} - \sqrt{\Delta_j})^2 \le \Delta_{ij} \le (\sqrt{\Delta_i} + \sqrt{\Delta_j})^2,
    \label{eq:pairwise_inequality}
\end{equation}
which must hold for every pair $i<j$.
For $n\ge 4$, however, the full positive semi-definiteness condition 
imposes additional correlations among the $\Delta_{ij}$ 
that are not captured by the pairwise bounds alone.
Concretely, all principal minors of $M$---not just the $2\times2$ ones---must be non-negative,
which translates into multi-index constraints on the $\Delta$ parameters.}

{Any successful matching of a Witten diagram must satisfy these full positivity conditions,
not merely the pairwise inequalities~\eqref{eq:pairwise_inequality}.
We will verify this explicitly for the matched examples in Section~\ref{sec:linear_comb},
where the chosen $\Delta_i$ and $\Delta_{ij}$ are shown to descend from a legitimate 
positive semi-definite covariance matrix.}\\

\textbf{Chord path integral: a brief review}

Before presenting our results for higher-point contact correlators, we briefly recall the chord path integral formalism developed in Refs.~\cite{Berkooz:2024evs,Berkooz:2024ofm} (see also Ref.~\cite{Jia:2025tvn} for the application to contact diagrams). Related path-integral treatments of chord diagrams have also been developed in the
study of the chaos to integrability transition in
DSSYK \cite{Gao:2024lve}. This technique converts the combinatorial sum over chord diagrams into a functional integral over a bilocal field, enabling the computation of correlation functions directly in the conformal limit.

The basic dynamical variable is the chord density $n(\tau_a,\tau_b)$, defined such that $n(\tau_a,\tau_b) d\tau_a d\tau_b$ counts the number of $H$-chords connecting intervals $[\tau_a,\tau_a+d\tau_a]$ and $[\tau_b,\tau_b+d\tau_b]$ on the Euclidean thermal circle. In the $q=e^{-2\lambda}\to 1^{-}$ limit, the partition function of the chord system is governed by the action \cite{Berkooz:2024evs,Berkooz:2024ofm}
\begin{align}
	S[n] = \frac{1}{4}\int_0^\beta d\tau_a\int_0^\beta d\tau_b\int_{\tau_a}^{\tau_b} d\tau_c\int_{\tau_b}^{\tau_a} d\tau_d\, n(\tau_a,\tau_b) n(\tau_c,\tau_d) \nonumber \\
	+ \frac{1}{2}\int_0^\beta d\tau_a\int_0^\beta d\tau_b\, n(\tau_a,\tau_b)\bigl[\log n(\tau_a,\tau_b)-1\bigr],
	\label{eq:chord_action}
\end{align}
with the periodicity convention $\int_{\tau_b}^{\tau_a} \equiv \int_{\tau_b}^{\tau_a+\beta}$ for $\tau_a<\tau_b$. The saddle-point equations are most conveniently expressed through the auxiliary field
\begin{equation}
	g(\tau_a,\tau_b) \coloneqq -\int_{\tau_a}^{\tau_b} d\tau \int_{\tau_b}^{\tau_a} d\tau'\, n(\tau,\tau')
	\quad\Longrightarrow\quad
	n(\tau_a,\tau_b) = -\frac{1}{2}\partial_{\tau_a}\partial_{\tau_b}\, g(\tau_a,\tau_b).
	\label{eq:g_def}
\end{equation}
At the saddle, $g$ satisfies 
\begin{eqnarray}
    \exp(g(\tau_a,\tau_b)) = \cos^2(\pi\nu/2)/\cos^2[\frac{\pi\nu}{2}(1-2|\tau_b-\tau_a|/\beta)],
\end{eqnarray}
which in the zero-temperature limit reduces to the conformal form
\begin{equation}
	\exp(g(\tau_a,\tau_b)) = \frac{1}{(\tau_b-\tau_a)^2}.
	\label{eq:g_zerot}
\end{equation}

Correlation functions of probe operators are obtained by inserting the appropriate intersection factors into the path integral. For a two-point function, the $O$-chord intersects $H$-chords whose endpoints lie in the intervals $(\tau_1,\tau_2)$, giving
\begin{equation}
	\langle O(\tau_2) O(\tau_1) \rangle = \Bigl\langle e^{-\Delta \int_{\tau_1}^{\tau_2} d\tau \int_{\tau_1}^{\tau_2} d\tau'\, n(\tau,\tau')} \Bigr\rangle
	= \langle e^{\Delta\, g(\tau_1,\tau_2)} \rangle,
	\label{eq:two_point_pathint}
\end{equation}
which at the saddle point yields $|\tau_{12}|^{-2\Delta}$, {where $\Delta$ is the conformal dimension of the probe operator defined in \eqref{eq:Delta_def}. For contact correlators with $n$ probe operators meeting at a point, the exponential weight involves multiple boundary integrals over the chord density, each associated with the intersection of $H$-chords with one or more probe legs.} Converting these integrals to the $g$ variables via \eqref{eq:g_def} and evaluating the saddle produces the factorized power-law forms quoted in Sections~\ref{subsec:3pt}-\ref{subsec:5pt6pt}. 

In the next section, we will apply the above theory to compute three-, four-, five- and six-point functions.

\section{Contact Correlators from Chord Diagrams and Matching with Witten Diagrams}
\label{sec:correlators}

In this section, we employ the chord diagram technique and the chord path integral formalism to compute explicit expressions for contact three-, four-, five-, and six-point correlation functions. We then systematically compare our results with known scalar contact Witten diagrams in AdS$_2$ \cite{Bliard:2022xsm}, and discuss the effects of lattice size and different configurations on the matching.

\subsection{Three-point functions}
\label{subsec:3pt}

	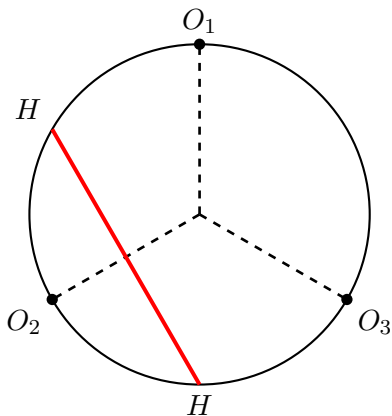
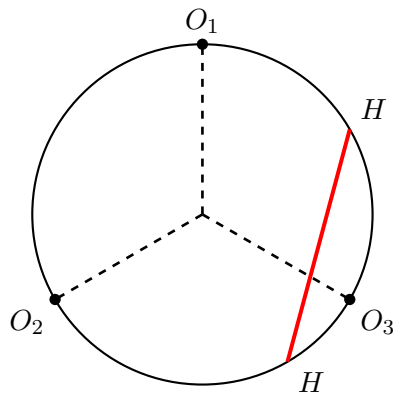
\begin{figure}[htbp]
		\centering
		\begin{subfigure}[c]{0.45\textwidth}  
			\centering
			\begin{tikzpicture}[
				thick,
				scale=0.9,
				dot/.style={circle, fill=black, inner sep=1.5pt},
				probe/.style={dashed, line width=1pt},
				Hchord/.style={line width=1.5pt, red}
				]
				
				\useasboundingbox (-3.2,-3.2) rectangle (3.2,3.2);
				
				\draw (0,0) circle (2.5cm);
				
				\foreach \i/\angle in {1/90, 2/210, 3/330}
				\node[dot] (p\i) at (\angle:2.5) {};
				
				\foreach \i/\angle in {1/90, 2/210, 3/330}
				\draw[probe] (0,0) -- (\angle:2.5);
				
				\node[above] at (p1) {$O_1$};
				\node[below left] at (p2) {$O_2$};
				\node[below right] at (p3) {$O_3$};
				
				\coordinate (Hstart) at (150:2.5);    
				\coordinate (Hend) at (270:2.5);      
				
				\draw[Hchord] (Hstart) -- (Hend);
				
				\node[above left] at (Hstart) {$H$};
				\node[below] at (Hend) {$H$};

			\end{tikzpicture}
			\caption{H-chord intersecting probe $O_2$}
			\label{fig:three-point-O2}
		\end{subfigure}
		\hfill
		\begin{subfigure}[c]{0.45\textwidth}  
			\centering
			\begin{tikzpicture}[
				thick,
				scale=0.9,
				dot/.style={circle, fill=black, inner sep=1.5pt},
				probe/.style={dashed, line width=1pt},
				Hchord/.style={line width=1.5pt, red}
				]
				
				\useasboundingbox (-3.2,-3.2) rectangle (3.2,3.2);
				
				\draw (0,0) circle (2.5cm);
				
				\foreach \i/\angle in {1/90, 2/210, 3/330}
				\node[dot] (p\i) at (\angle:2.5) {};
				
				\foreach \i/\angle in {1/90, 2/210, 3/330}
				\draw[probe] (0,0) -- (\angle:2.5);
				
				\node[above] at (p1) {$O_1$};
				\node[below left] at (p2) {$O_2$};
				\node[below right] at (p3) {$O_3$};
				
				\coordinate (Hstart) at (30:2.5);     
				\coordinate (Hend) at (300:2.5);      
				
				\draw[Hchord] (Hstart) -- (Hend);
				
				\node[above right] at (Hstart) {$H$};
				\node[below right] at (Hend) {$H$};

			\end{tikzpicture}
			\caption{H-chord intersecting probes $O_1$ and $O_3$}
			\label{fig:three-point-O1O3}
		\end{subfigure}
		
		\caption{Three-point contact chord diagrams. (a) A H-chord intersect probe $O_2$. (b) A  H-chord intersecting probe $O_3$. Dashed lines represent probe operators $O_i$, red lines represent H-chords. The overall factor for three-point contact diagrams is $1/\sqrt{N}$.}
		\label{fig:three-point-contact}
	\end{figure}

\par The three-point contact correlator, derived from the chord path integral as described in \cite{Jia:2025tvn}, takes a particularly simple form in the zero-temperature (conformal) limit. For operators $O_1, O_2, O_3$ with conformal dimensions $\Delta_1, \Delta_2, \Delta_3$ inserted at Euclidean times $\tau_1 < \tau_2 < \tau_3$, we obtain
\begin{equation}
    \langle O_3(\tau_3) O_2(\tau_2) O_1(\tau_1) \rangle_{\text{contact}} = \frac{C_{123}}{\sqrt{N}} \, \frac{1}{|\tau_{12}|^{\Delta_1+\Delta_2-\Delta_3} |\tau_{13}|^{\Delta_1+\Delta_3-\Delta_2} |\tau_{23}|^{\Delta_2+\Delta_3-\Delta_1}} ,
    \label{eq:3pt_result}
\end{equation}
where $\tau_{ij} = \tau_i - \tau_j$ and $C_{123}$ is a constant that depends on the microscopic details of the probe operators. 

We note that three-point contact contributions are only possible for odd lattice sizes $L$. For $L=3$, the allowed configurations are $(+,+,+)$ and $(-,-,-)$, which satisfy the closure condition $\sum_i \epsilon_i = 0 \mod 3$. For $L>3$, no additional configurations appear for three-point functions;  Thus, $L=3$ is the minimal and essentially unique choice for three-point contact diagrams in this framework.

The expression \eqref{eq:3pt_result} is precisely the form dictated by conformal symmetry for a three-point function in a one-dimensional CFT. This perfect agreement reinforces the interpretation of our chord contact diagrams as encoding bulk contact interactions.

\subsection{Four‑point contact functions}
\label{subsec:4pt}

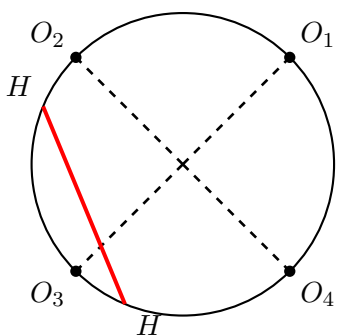
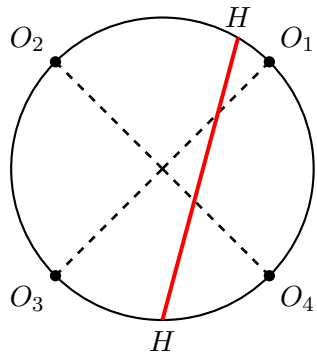
\begin{figure}[htbp]
	\centering
	\begin{subfigure}[t]{0.45\textwidth}
		\centering
		\begin{tikzpicture}[
			thick,
			scale=0.8,
			dot/.style={circle, fill=black, inner sep=1.5pt},
			probe/.style={dashed, line width=1pt},
			Hchord/.style={line width=1.5pt, red}
			]
			
			\draw (0,0) circle (2.5cm);
			
			\foreach \i/\angle in {1/45, 2/135, 3/225, 4/315}
			\node[dot] (p\i) at (\angle:2.5) {};
			
			\foreach \i/\angle in {1/45, 2/135, 3/225, 4/315}
			\draw[probe] (0,0) -- (\angle:2.5);
			
			\node[above right] at (p1) {$O_1$};
			\node[above left] at (p2) {$O_2$};
			\node[below left] at (p3) {$O_3$};
			\node[below right] at (p4) {$O_4$};
			
			\coordinate (Hstart) at (157.5:2.5);  
			\coordinate (Hend) at (247.5:2.5);    
			
			\draw[Hchord] (Hstart) -- (Hend);
			
			\node[above left] at (Hstart) {$H$};
			\node[below right] at (Hend) {$H$};

		\end{tikzpicture}
		\caption{H-chord intersecting one probe leg}
		\label{fig:four-point-single}
	\end{subfigure}
	\hfill
	\begin{subfigure}[t]{0.45\textwidth}
		\centering
		\begin{tikzpicture}[
			thick,
			scale=0.8,
			dot/.style={circle, fill=black, inner sep=1.5pt},
			probe/.style={dashed, line width=1pt},
			Hchord/.style={line width=1.5pt, red}
			]
			
			\draw (0,0) circle (2.5cm);
			
			\foreach \i/\angle in {1/45, 2/135, 3/225, 4/315}
			\node[dot] (p\i) at (\angle:2.5) {};
			
			\foreach \i/\angle in {1/45, 2/135, 3/225, 4/315}
			\draw[probe] (0,0) -- (\angle:2.5);
			
			\node[above right] at (p1) {$O_1$};
			\node[above left] at (p2) {$O_2$};
			\node[below left] at (p3) {$O_3$};
			\node[below right] at (p4) {$O_4$};
			
			\coordinate (Hstart) at (60:2.5);     
			\coordinate (Hend) at (270:2.5);      
			
			\draw[Hchord] (Hstart) -- (Hend);
			
			\node[above] at (Hstart) {$H$};
			\node[below] at (Hend) {$H$};
			
		\end{tikzpicture}
		\caption{H-chord intersecting two probe legs}
		\label{fig:four-point-double}
	\end{subfigure}
	
	\caption{Four-point contact chord diagrams. (a) A single H-chord intersects one probe leg. (b) A single H-chord intersects two probe legs. Dashed lines represent probe operators $O_i$, solid lines represent H-chords.}
	\label{fig:four-point-contact}
\end{figure}

We now turn to four-point functions, which exhibit a much richer structure. 
The allowed configurations depend crucially on the lattice size $L$:
\begin{itemize}
    \item For $L=2$, the condition $\sum_i \epsilon_i \equiv 0 \pmod 2$ requires an even number of $+1$'s. 
    With $n=4$, this gives all sequences with $0$, $2$, or $4$ plus signs: 
    $(+,+,+,+)$, $(-,-,-,-)$, and the six permutations with two $+$ and two $-$.
    Up to overall sign reversal, the independent classes are 
    $(+,+,+,+)$, $(+,+,-,-)$, $(+,-,+,-)$, and $(+,-,-,+)$.
    
    \item For $L=3$, the sum of four $\pm1$ values can only be $0$, $\pm2$, or $\pm4$.
    Divisibility by $3$ selects $\sum_i \epsilon_i = 0$, i.e.\ configurations with two $+$ and two $-$.
    All six such permutations are allowed, giving the independent classes 
    $(+,+,-,-)$, $(+,-,+,-)$, and $(+,-,-,+)$ (up to overall sign reversal).
    
    \item For $L=4$, the condition $\sum_i \epsilon_i \equiv 0 \pmod 4$ admits sums $0$, $\pm4$.
    Hence all two-positive-two-negative configurations (sum $0$) as well as the all-positive 
    $(+,+,+,+)$ and all-negative $(-,-,-,-)$ (sum $\pm4$) are allowed.
    The independent classes are $(+,+,+,+)$, $(+,+,-,-)$, $(+,-,+,-)$, and $(+,-,-,+)$.
    
    \item Larger $L$ does not provide new configurations for four points; 
    the set of possible sign sequences $\{\epsilon_i\}$ with $\sum_i \epsilon_i = 0 \mod L$ 
    and $\epsilon_i = \pm 1$ is exhausted by the above lists.
\end{itemize}

{To illustrate the general procedure with a concrete example, we provide a detailed computation of contact four-point functions on a size-$L=4$ periodic lattice---the simplest case among $L\ge3$ that admits the all-plus configuration $(+,+,+,+)$. 
The flux quantization condition for $L=4$ is $F = \pi n$, which is a special case of the general formula $F = 4\pi n/L$ derived in \eqref{eq:flux_quantization}; 
the algebraic structure of the chord rules is unchanged. 
Thus the results obtained here exemplify the method in full generality.} \\

\textbf{Representation matrices and periodisation}
\par For a periodic lattice of size $L=4$ we choose the following matrix representation of the Fock‑space oscillators:
\begin{align}
	\mathbb{S}^3 &= \frac12 \operatorname{diag}(3,\;1,\;-1,\;-3), \nonumber \\[2pt]
	P^+ &= \begin{pmatrix}
		0 & 1 & 0 & 0 \\
		0 & 0 & 1 & 0 \\
		0 & 0 & 0 & 1 \\
		1 & 0 & 0 & 0
	\end{pmatrix}, \qquad 
	P^- = (P^+)^\dagger .
\end{align}
The operators $P^\pm$ satisfy $(P^\pm)^4 = 1$ and implement periodic shifts on a four‑site chain. The magnetic operators are defined as
\begin{equation*}
	T_\mu^\pm = P_\mu^\pm \prod_{\rho\neq\mu} \exp\!\Bigl(\pm \frac{i}{2} F_{\mu\rho} \mathbb{S}_\rho^3\Bigr),
\end{equation*}
and similarly for the probe operators $\tilde T_\mu^{(i)\pm}$ with fluxes $\tilde F_{\mu\rho}^{(i)}$. The algebra \eqref{eq:magnetic algebra} holds provided the fluxes satisfy the rationality condition
\begin{equation}
	F_{\mu\rho} \in \pi\mathbb Z , \qquad \tilde F_{\mu\rho}^{(i)} \in \pi\mathbb Z ,
	\label{eq:rationality_L4}
\end{equation}
which guarantees the commutation relation $e^{i\frac{F}{2}\mathbb{S}^3}P^\pm e^{-i\frac{F}{2}\mathbb{S}^3}=e^{\pm i\frac{F}{2}}P^\pm$ for $F$ an integer multiple of $\pi$. \\

\textbf{Allowed configurations and closure condition.}

For a four-point contact insertion, all probe operators act on the same lattice direction $\mu$. A configuration is specified by a sign sequence $\{\epsilon_i=\pm1\}$ indicating whether $\tilde T_\mu^{(i)+}$ or $\tilde T_\mu^{(i)-}$ appears in the expansion. Closure of the path in the $\mu$-direction requires
\begin{equation}
    \sum_{i=1}^4 \epsilon_i = 0 \mod L,
    \label{eq:closure_general}
\end{equation}
as discussed in Section~\ref{sec:closure}. The complete classification of allowed configurations for $L=2,3,4$ is given in 
Section~\ref{sec:closure} and Table~\ref{tab:configs_L}; the $L=4$ case admits the all-plus, all-minus, and all two-positive-two-negative sequences.In what follows, we work on the $L=4$ lattice and construct the chord rules explicitly. \\

\textbf{Flux constraint to avoid exponential suppression}

{For a given configuration $\{\epsilon_i\}$, averaging over the fluxes from the directions $\rho\neq\mu$ produces the factor given in \eqref{eq:trace_factor}, which for $n=4$ takes the explicit form}
\begin{equation}
    \Bigl\langle \operatorname{Tr}\!\Bigl[ \exp\!\Bigl( \frac{i}{2}\sum_{i=1}^4 \epsilon_i \tilde F^{(i)} S^3 \Bigr) \Bigr] \Bigr\rangle^{N-1},
    \label{eq:trace_factor_L4}
\end{equation}
where the trace is taken over the four-dimensional representation of $S^3$. 
Using the eigenvalues of $S^3$, one finds
\begin{equation}
    \operatorname{Tr} e^{i\theta \mathbb{S}^3} = e^{3i\theta/2}+e^{i\theta/2}+e^{-i\theta/2}+e^{-3i\theta/2}
    = 2\cos\frac{\theta}{2}+2\cos\frac{3\theta}{2}.
    \label{eq:trace_S3_L4}
\end{equation}
This trace attains its maximal absolute value $4$ when $\theta = 0 \mod 2\pi$. 
Consequently, to avoid an exponential suppression of order 
$\bigl(|\operatorname{Tr} e^{i\theta S^3}|/4\bigr)^{N-1}$ in the large-$N$ limit, 
we must impose
\begin{equation}
    \frac12\sum_{i=1}^4 \epsilon_i \tilde F^{(i)} \in 2\pi\mathbb Z
    \quad\Longrightarrow\quad
    \sum_{i=1}^4 \epsilon_i \tilde F^{(i)} \in 4\pi\mathbb Z .
    \label{eq:flux_constraint_L4}
\end{equation}
For the uniform configuration $(+,+,+,+)$ this yields $\sum_i \tilde F^{(i)}\in 4\pi\mathbb Z$, 
while for the alternating configuration $(+,-,+,-)$ one recovers 
$\tilde F^{(1)}-\tilde F^{(2)}+\tilde F^{(3)}-\tilde F^{(4)}\in 4\pi\mathbb Z$.
{These conditions are the $L=4$ realization of the universal constraint in \eqref{eq:universal_constraint}, confirming that the $L=4$ lattice fits consistently into the general framework. }\\

\textbf{Chord rules and constraints for the size-4 lattice}

We now derive the chord diagram rules for the allowed configurations on the size-4 lattice. Since the contact point involves exactly four probe chords, each $H$-chord can intersect at most two probe chords. Consequently, only two-index joint factors $\Delta_{ij}$ appear, and there are no triple- or higher-order joint factors in four-point contact diagrams.
\begin{itemize}
	\item When an $H$-chord intersects a single probe chord $O_i$, the factor is
    \begin{eqnarray}
        q^{\Delta_i} = \big\langle \cos\big( \frac{F + \tilde{F}^{(i)}}{2} \big) \big\rangle.
    \end{eqnarray}
	\item When an $H$-chord intersects two probe chords $O_i$ and $O_j$, the factor is
	\begin{equation}
		q^{\Delta_{ij}} = 
		\begin{cases}
			\big\langle \cos\big( \frac{2F + \tilde{F}^{(i)} + \tilde{F}^{(j)}}{2} \big) \big\rangle, & \epsilon_i = \epsilon_j, \\[4pt]
			\big\langle \cos\big( \frac{\tilde{F}^{(i)} - \tilde{F}^{(j)}}{2} \big) \big\rangle, & \epsilon_i \neq \epsilon_j.
		\end{cases}
		\label{eq:Delta_ij_general}
	\end{equation}
\end{itemize}
The flux constraints derived in \eqref{eq:flux_constraint_L4} impose relations among the $\Delta_{ij}$. For any allowed configuration, one can show that
\begin{equation}
	\Delta_{12} = \Delta_{34}, \qquad \Delta_{23} = \Delta_{14}.
	\label{eq:symmetry_Delta}
\end{equation}
 These symmetries follow from the linear relations among the fluxes and the fact that the $\Delta_{ij}$ are quadratic in the fluxes (in the conformal limit). For instance, for the alternating configuration $(+,-,+,-)$, the constraint $\tilde{F}^{(1)} - \tilde{F}^{(2)} + \tilde{F}^{(3)} - \tilde{F}^{(4)} \in 4\pi\mathbb{Z}$ implies $\tilde{F}^{(1)} - \tilde{F}^{(2)} \approx -(\tilde{F}^{(3)} - \tilde{F}^{(4)})$ up to an integer multiple of $4\pi$, leading to $\Delta_{12} \approx \Delta_{34}$ when the fluctuations are small. A similar argument holds for the other pairs. \\

\textbf{Path-integral computation of the four-point correlator}

\begin{figure}[htbp]
	\centering
	\footnotesize 
	
	\begin{subfigure}[c]{0.32\textwidth}
		\centering
		\begin{tikzpicture}[
			thick,
			scale=0.6,
			dot/.style={circle, fill=black, inner sep=1pt},
			probe/.style={dashed, line width=0.8pt},
			arrow/.style={line width=1.2pt, -{Stealth[length=4pt]}, blue}
			]
			
			\draw (0,0) circle (2cm);
			
			\foreach \i/\angle in {1/45, 2/135, 3/225, 4/315}
			\node[dot] (p\i) at (\angle:2) {};
			
			\foreach \i/\angle in {1/45, 2/135, 3/225, 4/315}
			\draw[probe] (0,0) -- (\angle:2);
			
			\node[above right] at (p1) {$O_1$};
			\node[above left] at (p2) {$O_2$};
			\node[below left] at (p3) {$O_3$};
			\node[below right] at (p4) {$O_4$};
			
			\coordinate (mid12) at (90:2);    
			\coordinate (mid34) at (270:2);   
			
			\draw[arrow] (mid12) -- (mid34);
			
			
		\end{tikzpicture}
		\caption{Arrow from 1,2 to 3,4}
		\label{fig:arrow-12-34}
	\end{subfigure}
	\hfill
	\begin{subfigure}[c]{0.32\textwidth}
		\centering
		\begin{tikzpicture}[
			thick,
			scale=0.6,
			dot/.style={circle, fill=black, inner sep=1pt},
			probe/.style={dashed, line width=0.8pt},
			arrow/.style={line width=1.2pt, -{Stealth[length=4pt]}, blue}
			]
			
			\draw (0,0) circle (2cm);
			
			\foreach \i/\angle in {1/45, 2/135, 3/225, 4/315}
			\node[dot] (p\i) at (\angle:2) {};
			
			\foreach \i/\angle in {1/45, 2/135, 3/225, 4/315}
			\draw[probe] (0,0) -- (\angle:2);
			
			\node[above right] at (p1) {$O_1$};
			\node[above left] at (p2) {$O_2$};
			\node[below left] at (p3) {$O_3$};
			\node[below right] at (p4) {$O_4$};
			
			\coordinate (mid23) at (180:2);   
			\coordinate (mid34) at (270:2);   
			
			\draw[arrow] (mid23) -- (mid34);
			
			
		\end{tikzpicture}
		\caption{Arrow from 2,3 to 3,4}
		\label{fig:arrow-23-34}
	\end{subfigure}
	\hfill
	\begin{subfigure}[c]{0.32\textwidth}
		\centering
		\begin{tikzpicture}[
			thick,
			scale=0.6,
			dot/.style={circle, fill=black, inner sep=1pt},
			probe/.style={dashed, line width=0.8pt},
			arrow/.style={line width=1.2pt, -{Stealth[length=4pt]}, blue}
			]
			
			\draw (0,0) circle (2cm);
			
			\foreach \i/\angle in {1/45, 2/135, 3/225, 4/315}
			\node[dot] (p\i) at (\angle:2) {};
			
			\foreach \i/\angle in {1/45, 2/135, 3/225, 4/315}
			\draw[probe] (0,0) -- (\angle:2);
			
			\node[above right] at (p1) {$O_1$};
			\node[above left] at (p2) {$O_2$};
			\node[below left] at (p3) {$O_3$};
			\node[below right] at (p4) {$O_4$};
			
			\coordinate (mid34) at (270:2);   
			\coordinate (mid41) at (0:2);     
			
			\draw[arrow] (mid34) -- (mid41);
			
			
		\end{tikzpicture}
		\caption{Arrow from 3,4 to 4,1}
		\label{fig:arrow-34-41}
	\end{subfigure}
	
	\vspace{0.5cm} 
	
	\begin{subfigure}[c]{0.32\textwidth}
		\centering
		\begin{tikzpicture}[
			thick,
			scale=0.6,
			dot/.style={circle, fill=black, inner sep=1pt},
			probe/.style={dashed, line width=0.8pt},
			arrow/.style={line width=1.2pt, -{Stealth[length=4pt]}, blue}
			]
			
			\draw (0,0) circle (2cm);
			
			\foreach \i/\angle in {1/45, 2/135, 3/225, 4/315}
			\node[dot] (p\i) at (\angle:2) {};
			
			\foreach \i/\angle in {1/45, 2/135, 3/225, 4/315}
			\draw[probe] (0,0) -- (\angle:2);
			
			\node[above right] at (p1) {$O_1$};
			\node[above left] at (p2) {$O_2$};
			\node[below left] at (p3) {$O_3$};
			\node[below right] at (p4) {$O_4$};
			
			\coordinate (mid41) at (0:2);     
			\coordinate (mid12) at (90:2);    
			
			\draw[arrow] (mid41) -- (mid12);
			
			
		\end{tikzpicture}
		\caption{Arrow from 4,1 to 1,2}
		\label{fig:arrow-41-12}
	\end{subfigure}
	\hfill
	\begin{subfigure}[c]{0.32\textwidth}
		\centering
		\begin{tikzpicture}[
			thick,
			scale=0.6,
			dot/.style={circle, fill=black, inner sep=1pt},
			probe/.style={dashed, line width=0.8pt},
			arrow/.style={line width=1.2pt, -{Stealth[length=4pt]}, blue}
			]
			
			\draw (0,0) circle (2cm);
			
			\foreach \i/\angle in {1/45, 2/135, 3/225, 4/315}
			\node[dot] (p\i) at (\angle:2) {};
			
			\foreach \i/\angle in {1/45, 2/135, 3/225, 4/315}
			\draw[probe] (0,0) -- (\angle:2);
			
			\node[above right] at (p1) {$O_1$};
			\node[above left] at (p2) {$O_2$};
			\node[below left] at (p3) {$O_3$};
			\node[below right] at (p4) {$O_4$};
			
			\coordinate (mid12) at (90:2);    
			\coordinate (mid34) at (270:2);   
			
			\draw[arrow] (mid12) -- (mid34);
			
			
		\end{tikzpicture}
		\caption{Arrow from 1,2 to 3,4}
		\label{fig:arrow-12-34-2}
	\end{subfigure}
	\hfill
	\begin{subfigure}[c]{0.32\textwidth}
		\centering
		\begin{tikzpicture}[
			thick,
			scale=0.6,
			dot/.style={circle, fill=black, inner sep=1pt},
			probe/.style={dashed, line width=0.8pt},
			arrow/.style={line width=1.2pt, -{Stealth[length=4pt]}, blue}
			]
			
			\draw (0,0) circle (2cm);
			
			\foreach \i/\angle in {1/45, 2/135, 3/225, 4/315}
			\node[dot] (p\i) at (\angle:2) {};
			
			\foreach \i/\angle in {1/45, 2/135, 3/225, 4/315}
			\draw[probe] (0,0) -- (\angle:2);
			
			\node[above right] at (p1) {$O_1$};
			\node[above left] at (p2) {$O_2$};
			\node[below left] at (p3) {$O_3$};
			\node[below right] at (p4) {$O_4$};
			
			\coordinate (mid41) at (0:2);     
			\coordinate (mid23) at (180:2);   
			
			\draw[arrow] (mid41) -- (mid23);
			
			
		\end{tikzpicture}
		\caption{Arrow from 4,1 to 2,3}
		\label{fig:arrow-41-23}
	\end{subfigure}
	
	\caption{converting $n(\tau,\tau')$ to $g(\tau,\tau')$ in the path integral four-point functions.}
	\label{fig:six-arrow-diagrams}
\end{figure}
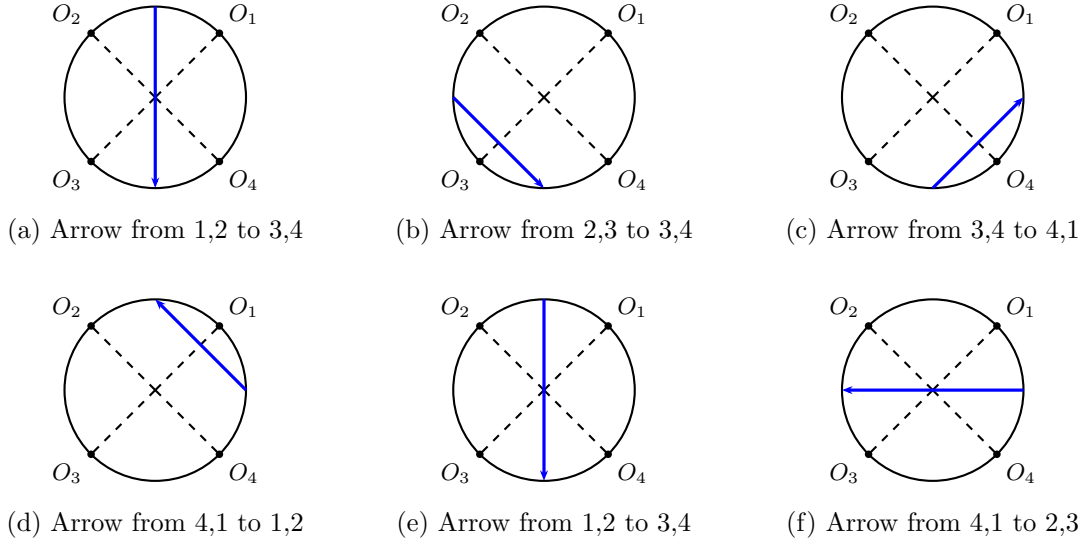

We now compute the Euclidean four-point contact correlator
\(\langle O_4(\tau_4)  O_3(\tau_3) O_2(\tau_2) O_1(\tau_1) \rangle_{\text{contact}}\) using the chord path integral.
Assuming the ordering \(\tau_4 >\tau_3 > \tau_2 > \tau_1\), the result takes the form
\begin{equation}
	\langle  O_4 O_3 O_2 O_1 \rangle_{\text{contact}} = \frac{1}{N} \big\langle e^{-K[n]} \big\rangle,
	\label{eq:4pt-correlator-generic}
\end{equation}
where the functional \(K[n]\) is given by
\begin{align}
K[n]  = & \frac{\Delta_{12}-\Delta_{1}-\Delta_{2}}{2} g_{12}+\frac{\Delta_{2}+\Delta_{4}-\Delta_{12}-\Delta_{23}}{2} g_{13} 
	 +\frac{\Delta_{23}-\Delta_{1}-\Delta_{4}}{2} g_{14}  \nonumber \\
	 &+\frac{\Delta_{23}-\Delta_{2}-\Delta_{3}}{2} g_{23} 
	 +\frac{\Delta_{1}+\Delta_{3}-\Delta_{12}-\Delta_{23}}{2} g_{24}+\frac{\Delta_{12}-\Delta_{3}-\Delta_{4}}{2} g_{34}.
	\label{eq:K-4pt}
\end{align}
{For completeness, we illustrate the conversion from $n(\tau,\tau')$ to $g(\tau,\tau')$ for both the three-point and four-point contact diagrams in Figs.~\ref{fig:three-arrow-diagrams} and~\ref{fig:six-arrow-diagrams}. Although the three-point calculation is omitted from the main text (the result is quoted in Section~\ref{subsec:3pt}), the diagrammatic procedure shown in Fig.~\ref{fig:three-arrow-diagrams} is structurally identical to the one used here for the four-point case.}

\begin{figure}[htbp]
	\centering
	
	\begin{subfigure}[c]{0.32\textwidth}
		\centering
		\begin{tikzpicture}[
			thick,
			scale=0.7,
			dot/.style={circle, fill=black, inner sep=1.2pt},
			probe/.style={dashed, line width=0.8pt},
			arrow/.style={line width=1.2pt, -{Stealth[length=4pt]}, blue}
			]
			
			\draw (0,0) circle (2.2cm);
			
			\foreach \i/\angle in {1/90, 2/210, 3/330}
			\node[dot] (p\i) at (\angle:2.2) {};
			
			\foreach \i/\angle in {1/90, 2/210, 3/330}
			\draw[probe] (0,0) -- (\angle:2.2);
			
			\node[above] at (p1) {$O_1$};
			\node[below left] at (p2) {$O_2$};
			\node[below right] at (p3) {$O_3$};
			
			\coordinate (mid31) at (30:2.2);   
			\coordinate (mid12) at (150:2.2);  
			
			\draw[arrow] (mid31) -- (mid12);
			
			
		\end{tikzpicture}
		\caption{Arrow from $3,1$ to $1,2$}
		\label{fig:arrow-31-12}
	\end{subfigure}
	\hfill
	\begin{subfigure}[c]{0.32\textwidth}
		\centering
		\begin{tikzpicture}[
			thick,
			scale=0.7,
			dot/.style={circle, fill=black, inner sep=1.2pt},
			probe/.style={dashed, line width=0.8pt},
			arrow/.style={line width=1.2pt, -{Stealth[length=4pt]}, blue}
			]
			
			\draw (0,0) circle (2.2cm);
			
			\foreach \i/\angle in {1/90, 2/210, 3/330}
			\node[dot] (p\i) at (\angle:2.2) {};
			
			\foreach \i/\angle in {1/90, 2/210, 3/330}
			\draw[probe] (0,0) -- (\angle:2.2);
			
			\node[above] at (p1) {$O_1$};
			\node[below left] at (p2) {$O_2$};
			\node[below right] at (p3) {$O_3$};
			
			\coordinate (mid12) at (150:2.2);  
			\coordinate (mid23) at (270:2.2);  
			
			\draw[arrow] (mid12) -- (mid23);
			
			
		\end{tikzpicture}
		\caption{Arrow from $1,2$ to $2,3$}
		\label{fig:arrow-12-23}
	\end{subfigure}
	\hfill
	\begin{subfigure}[c]{0.32\textwidth}
		\centering
		\begin{tikzpicture}[
			thick,
			scale=0.7,
			dot/.style={circle, fill=black, inner sep=1.2pt},
			probe/.style={dashed, line width=0.8pt},
			arrow/.style={line width=1.2pt, -{Stealth[length=4pt]}, blue}
			]
			
			\draw (0,0) circle (2.2cm);
			
			\foreach \i/\angle in {1/90, 2/210, 3/330}
			\node[dot] (p\i) at (\angle:2.2) {};
			
			\foreach \i/\angle in {1/90, 2/210, 3/330}
			\draw[probe] (0,0) -- (\angle:2.2);
			
			\node[above] at (p1) {$O_1$};
			\node[below left] at (p2) {$O_2$};
			\node[below right] at (p3) {$O_3$};
			
			\coordinate (mid23) at (270:2.2);  
			\coordinate (mid31) at (30:2.2);   
			
			\draw[arrow] (mid23) -- (mid31);
			
			
		\end{tikzpicture}
		\caption{Arrow from $2,3$ to $3,1$}
		\label{fig:arrow-23-31}
	\end{subfigure}
	
	\caption{Converting $n(\tau,\tau')$ to $g(\tau,\tau')$n the path integral three-point functions.}
	\label{fig:three-arrow-diagrams}
\end{figure}

\textbf{Result in the zero-temperature limit}

Evaluating the path integral in the saddle-point approximation and taking the zero-temperature limit where $g_{ij} = -\log |\tau_{ij}|^2$, we obtain the explicit four-point contact correlator:
\begin{equation}
    \langle O_4 O_3 O_2 O_1 \rangle_{\mathcal{C}} = \frac{1}{N} \prod_{1 \le i < j \le 4} \frac{1}{|\tau_{ij}|^{\gamma_{ij}^{(\mathcal{C})}}},
    \label{eq:4pt-explicit}
\end{equation}
where the exponents $\gamma_{ij}^{(\mathcal{C})}$ are read off from the functional $K[n]$ . 
{For any allowed four-point configuration, the flux constraint~\eqref{eq:flux_constraint_L4} implies
$\Delta_{34}=\Delta_{12}$ and $\Delta_{14}=\Delta_{23}$, and the non‑vanishing exponents are
\begin{equation}
\begin{aligned}
    \gamma_{12} &= \Delta_{12} - \Delta_1 - \Delta_2, \qquad 
    \gamma_{23} = \Delta_{23} - \Delta_2 - \Delta_3, \\
    \gamma_{34} &= \Delta_{12} - \Delta_3 - \Delta_4, \qquad 
    \gamma_{14} = \Delta_{23} - \Delta_1 - \Delta_4, \\
    \gamma_{13} &= \Delta_2 + \Delta_4 - \Delta_{12} - \Delta_{23}, \qquad
    \gamma_{24} = \Delta_1 + \Delta_3 - \Delta_{12} - \Delta_{23},
\end{aligned}
\label{eq:gamma_4pt}
\end{equation}
Notice that $\gamma_{13}$ and $\gamma_{24}$ are not of the simple form $\Delta_i+\Delta_j-\Delta_{ij}$; 
their specific structure follows from the detailed chord intersections encoded in $K[n]$.}
{To exhibit the dependence on the independent cross-ratio, we follow Ref.~\cite{Jia:2025tvn} and set}
\begin{equation}
    \tau_1 = 0,\quad \tau_2 = x,\quad \tau_3 = 1,\quad \tau_4 \to \infty,
    \label{eq:tau_4pt}
\end{equation}
and define the scaled correlator
\begin{equation}
    I(x) \coloneqq \lim_{\tau_4 \to \infty} |\tau_4|^{2\Delta_4} \,
    \langle O_4 O_3 O_2 O_1 \rangle_{\mathrm{contact}} .
    \label{eq:I_4pt}
\end{equation}
{This quantity depends only on the cross‑ratio $x$ and can be directly compared with the AdS$_2$ contact Witten diagrams catalogued in Ref.~\cite{Bliard:2022xsm}.}
After extracting the overall scaling by sending $\tau_4 \to \infty$, the correlator reduces to a function of the cross‑ratio $x = \tau_{12}\tau_{34}/(\tau_{13}\tau_{24})$.
Then
\begin{align}
	I(x) = \frac{1}{N} x^{\Delta_{12}-\Delta_1-\Delta_2} (1-x)^{\Delta_{23}-\Delta_2-\Delta_3}.
	\label{eq:4pt-cross-ratios}
\end{align}
This expression manifests the expected factorized power-law form dictated by conformal symmetry for a four-point contact diagram. We can freely choose the crossing conformal factors while maintaining the positivity constraints, in order to match the interactions in the bulk.

For a given configuration $\mathcal{C}$, the chord path integral yields a pure power‑law form dictated by conformal symmetry:
\begin{equation}
    I_{\mathcal{C}}(x) = \frac{1}{N} \, x^{a_{\mathcal{C}}} (1-x)^{b_{\mathcal{C}}},
    \label{eq:4pt_pure_power}
\end{equation}
where the exponents $a_{\mathcal{C}}, b_{\mathcal{C}}$ are determined by the flux parameters $\Delta_i^{(\mathcal{C})}$ and $\Delta_{ij}^{(\mathcal{C})}$ .

However, the AdS$_2$ contact Witten diagrams catalogued in~\cite{Bliard:2022xsm} are not all pure power laws; many contain logarithmic singularities (e.g., $x^a \log x$ or $(1-x)^b \log(1-x)$) or rational functions that are sums of several power laws.

\subsection{Five-point and six-point functions}
\label{subsec:5pt6pt}

	\begin{figure}[htbp]
		\centering
		\begin{subfigure}[c]{0.45\textwidth}
			\centering
			\begin{tikzpicture}[
				thick,
				scale=0.8,
				dot/.style={circle, fill=black, inner sep=1.5pt},
				probe/.style={dashed, line width=1pt},
				Hchord/.style={line width=1.5pt, red}
				]
				
				\useasboundingbox (-3.2,-3.2) rectangle (3.2,3.2);
				
				\draw (0,0) circle (2.5cm);
				
				\foreach \i/\angle in {1/90, 2/162, 3/234, 4/306, 5/18}
				\node[dot] (p\i) at (\angle:2.5) {};
				
				\foreach \i/\angle in {1/90, 2/162, 3/234, 4/306, 5/18}
				\draw[probe] (0,0) -- (\angle:2.5);
				
				\node[above] at (p1) {$O_1$};
				\node[above left] at (p2) {$O_2$};
				\node[below left] at (p3) {$O_3$};
				\node[below right] at (p4) {$O_4$};
				\node[above right] at (p5) {$O_5$};
				
				\coordinate (Hstart) at (198:2.5);    
				\coordinate (Hend) at (270:2.5);      
				
				\draw[Hchord] (Hstart) -- (Hend);
				
				\node[above left] at (Hstart) {$H$};
				\node[below] at (Hend) {$H$};

			\end{tikzpicture}
			\caption{H-chord intersecting $O_3$}
			\label{fig:five-point-single}
		\end{subfigure}
		\hfill
		\begin{subfigure}[c]{0.45\textwidth}
			\centering
			\begin{tikzpicture}[
				thick,
				scale=0.8,
				dot/.style={circle, fill=black, inner sep=1.5pt},
				probe/.style={dashed, line width=1pt},
				Hchord/.style={line width=1.5pt, red}
				]
				
				\useasboundingbox (-3.2,-3.2) rectangle (3.2,3.2);
				
				\draw (0,0) circle (2.5cm);
				
				\foreach \i/\angle in {1/90, 2/162, 3/234, 4/306, 5/18}
				\node[dot] (p\i) at (\angle:2.5) {};
				
				\foreach \i/\angle in {1/90, 2/162, 3/234, 4/306, 5/18}
				\draw[probe] (0,0) -- (\angle:2.5);
				
				\node[above] at (p1) {$O_1$};
				\node[above left] at (p2) {$O_2$};
				\node[below left] at (p3) {$O_3$};
				\node[below right] at (p4) {$O_4$};
				\node[above right] at (p5) {$O_5$};
				
				\coordinate (Hstart) at (126:2.5);    
				\coordinate (Hend) at (342:2.5);      
				
				\draw[Hchord] (Hstart) -- (Hend);
				
				\node[above left] at (Hstart) {$H$};
				\node[above right] at (Hend) {$H$};

			\end{tikzpicture}
			\caption{H-chord intersecting $O_1$ and $O_5$}
			\label{fig:five-point-double}
		\end{subfigure}
		
		\caption{Five-point contact chord diagrams. (a) A H-chord intersecting probe $O_3$. (b) A  H-chord intersecting probes $O_1$ and $O_5$. Dashed lines represent probe operators $O_i$, red lines represent H-chords.}
		\label{fig:five-point-contact}
	\end{figure}
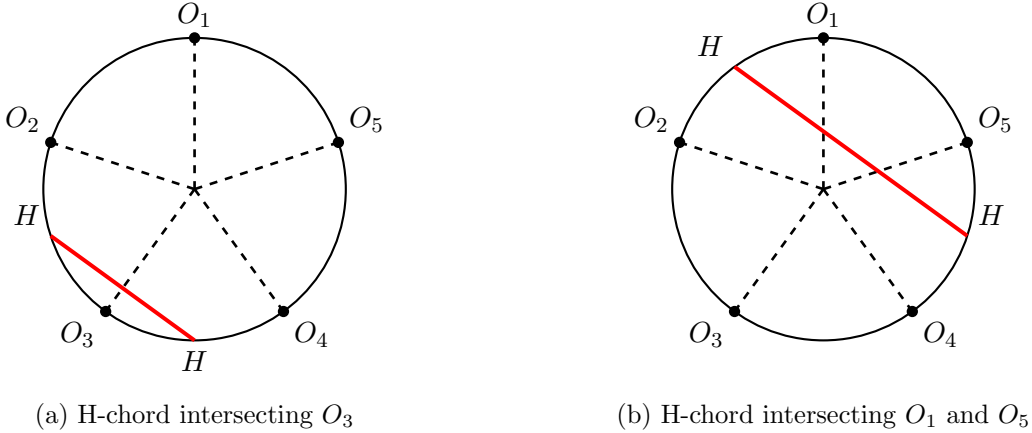

In this subsection, we provide a detailed exposition of the construction and computation of five-point contact correlation functions within the generalized chord-diagram framework. The increased complexity—arising from two independent cross-ratios and a richer set of allowed probe configurations—offers a stringent test of the method's generality.  \\

\textbf{Allowed configurations for five points}

For a periodic lattice of size \(L\), the closure condition for five probe operators reads
\begin{equation}
	\sum_{i=1}^{5} \epsilon_i = 0 \mod L, \qquad \epsilon_i = \pm 1 .
	\label{eq:5pt-closure}
\end{equation}
The set of solutions grows with \(L\). For the smallest odd lattices, which are required to admit non-trivial closed loops, we have:

\begin{itemize}
	\item \(L = 3\): The only allowed configurations are those with four plus signs and one minus sign (or vice versa), e.g., \((+,+,+,+,-)\). The net displacement is \(\pm 3\), which is a multiple of 3.
	\item \(L = 5\): The all-plus configuration \((+,+,+,+,+)\) and the all-minus configuration \((-,-,-,-,-)\) are allowed, since \(5 \equiv 0 \mod 5\).
	\item Larger odd \(L\) does not yield new configurations for five points, because the sum of five \(\pm 1\) values cannot reach a multiple of \(L\) unless \(L\) divides 3 or 5.
\end{itemize}

In what follows, we focus on the simplest non-trivial case \(L=3\), which already captures the essential features. \\

\textbf{Flux constraints and chord rules}

Working with the three-dimensional representation of the SU(2) algebra and imposing periodic boundary conditions, the magnetic fluxes are quantized as
\begin{equation}
	F = \frac{4n\pi}{3}, \quad n \in \mathbb{Z}.
	\label{eq:flux-quantization-5pt}
\end{equation}
The normalized Hamiltonian is taken to be
\begin{equation}
	H = \frac{1}{\sqrt{N}} \sum_{\mu=1}^{N} D_{\mu}, \qquad D_{\mu} = T_{\mu}^{+} + T_{\mu}^{-},
	\label{eq:H-5pt}
\end{equation}
and similarly for the probe operators \(O_i\) with their own fluxes \(\tilde{F}^{(i)}\).

The contact contribution to the moments is given by
\begin{equation}
	M^{\text{contact}}_{k_1,\dots,k_5} = \frac{1}{N^{2 + \sum_i k_i/2}} \sum_{\mu}
	\Big\langle \operatorname{Tr}\big( H^{k_5} \tilde{D}_{\mu}^{(5)} H^{k_4} \tilde{D}_{\mu}^{(4)} H^{k_3} \tilde{D}_{\mu}^{(3)} H^{k_2} \tilde{D}_{\mu}^{(2)} H^{k_1} \tilde{D}_{\mu}^{(1)} \big) \Big\rangle,
	\label{eq:5pt-moments}
\end{equation}
where the trace enforces the selection of closed loops in the Fock-space lattice. Expanding each \(\tilde{D}_{\mu}^{(i)}\) into \(\tilde{T}_{\mu}^{(i)\pm}\), the contributing terms satisfy
\begin{equation}
	\sum_{i=1}^{5} \eta_i = 0 \mod 3, \qquad \eta_i = \pm 1,
	\label{eq:sign-sum}
\end{equation}
and the average over fluxes yields a factor
\begin{equation}
	\Big\langle \cos\!\Big( \frac{1}{2} \sum_{i=1}^{5} \eta_i \tilde{F}^{(i)} \Big) \Big\rangle^{N-1}.
	\label{eq:flux-average-5pt}
\end{equation}
To avoid exponential suppression in the large-\(N\) limit, we impose the constraint
\begin{equation}
	\sum_{i=1}^{5} \eta_i \tilde{F}^{(i)} \in 4\pi \mathbb{Z}.
	\label{eq:flux-constraint-5pt}
\end{equation}

The weight of a chord diagram is then governed by the following rules:
\begin{itemize}
	\item Each intersection between an \(H\)-chord and a single probe chord \(O_i\) contributes a factor \(q^{\Delta_i}\), where
	\begin{equation}
		q^{\Delta_i} = \Big\langle \cos\!\Big( \frac{F + \tilde{F}^{(i)}}{2} \Big) \Big\rangle.
		\label{eq:Delta-i-5pt}
	\end{equation}
	\item Each intersection of an \(H\)-chord with two probe chords \(O_i\) and \(O_j\) contributes a factor \(q^{\Delta_{ij}}\), where
	\begin{equation}
		q^{\Delta_{ij}} = 
		\begin{cases}
			\big\langle \cos\!\big( F + \frac{\tilde{F}^{(i)} + \tilde{F}^{(j)}}{2} \big) \big\rangle, & \epsilon_i = \epsilon_j, \\[6pt]
			\big\langle \cos\!\big( \frac{\tilde{F}^{(i)} - \tilde{F}^{(j)}}{2} \big) \big\rangle, & \epsilon_i \neq \epsilon_j .
		\end{cases}
		\label{eq:Delta-ij-5pt}
	\end{equation}
	Here \(\epsilon_i\) denotes the sign of the corresponding \(\tilde{T}_{\mu}^{(i)\epsilon_i}\) operator in the chosen configuration.
\end{itemize}
In the conformal limit, these reduce to the quadratic expressions
\begin{equation}
	\Delta_i \approx \frac{\langle (F + \tilde{F}^{(i)})^2 \rangle}{4\langle F^2 \rangle}, \qquad
	\Delta_{ij} \approx 
	\begin{cases}
		\displaystyle \frac{\langle (F + (\tilde{F}^{(i)} + \tilde{F}^{(j)})/2)^2 \rangle}{4\langle F^2 \rangle}, & \epsilon_i = \epsilon_j, \\[10pt]
		\displaystyle \frac{\langle (\tilde{F}^{(i)} - \tilde{F}^{(j)})^2 \rangle}{4\langle F^2 \rangle}, & \epsilon_i \neq \epsilon_j .
	\end{cases}
	\label{eq:Delta-conformal-5pt}
\end{equation}

\textbf{Path-integral computation of the five-point correlator}

We now compute the Euclidean five-point contact correlator
\(\langle O_5(\tau_5) O_4(\tau_4) O_3(\tau_3) O_2(\tau_2) O_1(\tau_1) \rangle_{\text{contact}}\) using the chord path integral.
Assuming the ordering \(\tau_5 > \tau_4 > \tau_3 > \tau_2 > \tau_1\), the result takes the form
\begin{equation}
	\langle O_5 O_4 O_3 O_2 O_1 \rangle_{\text{contact}} = \frac{1}{N} \big\langle e^{-K[n]} \big\rangle,
	\label{eq:5pt-correlator-generic}
\end{equation}
where the functional \(K[n]\) is given by
\begin{align}
	K[n] ={}& \Delta_1 \int_{\tau_1}^{\tau_5} d\tau \int_{\tau_2}^{\tau_1} d\tau' \, n(\tau,\tau')
	+ \Delta_2 \int_{\tau_2}^{\tau_1} d\tau \int_{\tau_3}^{\tau_2} d\tau' \, n(\tau,\tau') \nonumber \\
	&+ \Delta_3 \int_{\tau_3}^{\tau_2} d\tau \int_{\tau_4}^{\tau_3} d\tau' \, n(\tau,\tau')
	+ \Delta_4 \int_{\tau_4}^{\tau_3} d\tau \int_{\tau_5}^{\tau_4} d\tau' \, n(\tau,\tau') \nonumber \\
	&+ \Delta_5 \int_{\tau_5}^{\tau_4} d\tau \int_{\tau_1}^{\tau_5} d\tau' \, n(\tau,\tau')
	+ \Delta_{12} \int_{\tau_3}^{\tau_2} d\tau \int_{\tau_1}^{\tau_5} d\tau' \, n(\tau,\tau') \nonumber \\
	&+ \Delta_{23} \int_{\tau_2}^{\tau_1} d\tau \int_{\tau_4}^{\tau_3} d\tau' \, n(\tau,\tau')
	+ \Delta_{34} \int_{\tau_3}^{\tau_2} d\tau \int_{\tau_5}^{\tau_4} d\tau' \, n(\tau,\tau') \nonumber \\
	&+ \Delta_{45} \int_{\tau_4}^{\tau_3} d\tau \int_{\tau_1}^{\tau_5} d\tau' \, n(\tau,\tau')
	+ \Delta_{15} \int_{\tau_2}^{\tau_1} d\tau \int_{\tau_5}^{\tau_4} d\tau' \, n(\tau,\tau').
	\label{eq:K-5pt}
\end{align}

\begin{figure}[htbp]
	\centering
	\scriptsize 
	
	\begin{subfigure}[c]{0.18\textwidth}
		\centering
		\begin{tikzpicture}[
			thick,
			scale=0.45,
			dot/.style={circle, fill=black, inner sep=0.8pt},
			probe/.style={dashed, line width=0.6pt},
			arrow/.style={line width=1pt, -{Stealth[length=3pt]}, blue}
			]
			
			\draw (0,0) circle (1.8cm);
			
			\foreach \i/\angle in {1/90, 2/162, 3/234, 4/306, 5/18}
			\node[dot] (p\i) at (\angle:1.8) {};
			
			\foreach \i/\angle in {1/90, 2/162, 3/234, 4/306, 5/18}
			\draw[probe] (0,0) -- (\angle:1.8);
			
			\node[above] at (p1) {\tiny $O_1$};
			\node[above left] at (p2) {\tiny $O_2$};
			\node[below left] at (p3) {\tiny $O_3$};
			\node[below right] at (p4) {\tiny $O_4$};
			\node[above right] at (p5) {\tiny $O_5$};
			
			\coordinate (mid51) at (54:1.8);   
			\coordinate (mid12) at (126:1.8);  
			
			\draw[arrow] (mid51) -- (mid12);
			
			
		\end{tikzpicture}
		\caption{$51\to12$}
		\label{fig:arrow-51-12}
	\end{subfigure}
	\hfill
	\begin{subfigure}[c]{0.18\textwidth}
		\centering
		\begin{tikzpicture}[
			thick,
			scale=0.45,
			dot/.style={circle, fill=black, inner sep=0.8pt},
			probe/.style={dashed, line width=0.6pt},
			arrow/.style={line width=1pt, -{Stealth[length=3pt]}, blue}
			]
			
			\draw (0,0) circle (1.8cm);
			
			\foreach \i/\angle in {1/90, 2/162, 3/234, 4/306, 5/18}
			\node[dot] (p\i) at (\angle:1.8) {};
			
			\foreach \i/\angle in {1/90, 2/162, 3/234, 4/306, 5/18}
			\draw[probe] (0,0) -- (\angle:1.8);
			
			\node[above] at (p1) {\tiny $O_1$};
			\node[above left] at (p2) {\tiny $O_2$};
			\node[below left] at (p3) {\tiny $O_3$};
			\node[below right] at (p4) {\tiny $O_4$};
			\node[above right] at (p5) {\tiny $O_5$};
			
			\coordinate (mid12) at (126:1.8);  
			\coordinate (mid23) at (198:1.8);  
			
			\draw[arrow] (mid12) -- (mid23);
			
			
		\end{tikzpicture}
		\caption{$12\to23$}
		\label{fig:arrow-12-23}
	\end{subfigure}
	\hfill
	\begin{subfigure}[c]{0.18\textwidth}
		\centering
		\begin{tikzpicture}[
			thick,
			scale=0.45,
			dot/.style={circle, fill=black, inner sep=0.8pt},
			probe/.style={dashed, line width=0.6pt},
			arrow/.style={line width=1pt, -{Stealth[length=3pt]}, blue}
			]
			
			\draw (0,0) circle (1.8cm);
			
			\foreach \i/\angle in {1/90, 2/162, 3/234, 4/306, 5/18}
			\node[dot] (p\i) at (\angle:1.8) {};
			
			\foreach \i/\angle in {1/90, 2/162, 3/234, 4/306, 5/18}
			\draw[probe] (0,0) -- (\angle:1.8);
			
			\node[above] at (p1) {\tiny $O_1$};
			\node[above left] at (p2) {\tiny $O_2$};
			\node[below left] at (p3) {\tiny $O_3$};
			\node[below right] at (p4) {\tiny $O_4$};
			\node[above right] at (p5) {\tiny $O_5$};
			
			\coordinate (mid23) at (198:1.8);  
			\coordinate (mid34) at (270:1.8);  
			
			\draw[arrow] (mid23) -- (mid34);
			
			
		\end{tikzpicture}
		\caption{$23\to34$}
		\label{fig:arrow-23-34}
	\end{subfigure}
	\hfill
	\begin{subfigure}[c]{0.18\textwidth}
		\centering
		\begin{tikzpicture}[
			thick,
			scale=0.45,
			dot/.style={circle, fill=black, inner sep=0.8pt},
			probe/.style={dashed, line width=0.6pt},
			arrow/.style={line width=1pt, -{Stealth[length=3pt]}, blue}
			]
			
			\draw (0,0) circle (1.8cm);
			
			\foreach \i/\angle in {1/90, 2/162, 3/234, 4/306, 5/18}
			\node[dot] (p\i) at (\angle:1.8) {};
			
			\foreach \i/\angle in {1/90, 2/162, 3/234, 4/306, 5/18}
			\draw[probe] (0,0) -- (\angle:1.8);
			
			\node[above] at (p1) {\tiny $O_1$};
			\node[above left] at (p2) {\tiny $O_2$};
			\node[below left] at (p3) {\tiny $O_3$};
			\node[below right] at (p4) {\tiny $O_4$};
			\node[above right] at (p5) {\tiny $O_5$};
			
			\coordinate (mid34) at (270:1.8);  
			\coordinate (mid45) at (342:1.8);  
			
			\draw[arrow] (mid34) -- (mid45);
			
			
		\end{tikzpicture}
		\caption{$34\to45$}
		\label{fig:arrow-34-45}
	\end{subfigure}
	\hfill
	\begin{subfigure}[c]{0.18\textwidth}
		\centering
		\begin{tikzpicture}[
			thick,
			scale=0.45,
			dot/.style={circle, fill=black, inner sep=0.8pt},
			probe/.style={dashed, line width=0.6pt},
			arrow/.style={line width=1pt, -{Stealth[length=3pt]}, blue}
			]
			
			\draw (0,0) circle (1.8cm);
			
			\foreach \i/\angle in {1/90, 2/162, 3/234, 4/306, 5/18}
			\node[dot] (p\i) at (\angle:1.8) {};
			
			\foreach \i/\angle in {1/90, 2/162, 3/234, 4/306, 5/18}
			\draw[probe] (0,0) -- (\angle:1.8);
			
			\node[above] at (p1) {\tiny $O_1$};
			\node[above left] at (p2) {\tiny $O_2$};
			\node[below left] at (p3) {\tiny $O_3$};
			\node[below right] at (p4) {\tiny $O_4$};
			\node[above right] at (p5) {\tiny $O_5$};
			
			\coordinate (mid45) at (342:1.8);  
			\coordinate (mid51) at (54:1.8);   
			
			\draw[arrow] (mid45) -- (mid51);
			
			
		\end{tikzpicture}
		\caption{$45\to51$}
		\label{fig:arrow-45-51}
	\end{subfigure}
	
	\vspace{0.3cm}
	
	\begin{subfigure}[c]{0.18\textwidth}
		\centering
		\begin{tikzpicture}[
			thick,
			scale=0.45,
			dot/.style={circle, fill=black, inner sep=0.8pt},
			probe/.style={dashed, line width=0.6pt},
			arrow/.style={line width=1pt, -{Stealth[length=3pt]}, blue}
			]
			
			\draw (0,0) circle (1.8cm);
			
			\foreach \i/\angle in {1/90, 2/162, 3/234, 4/306, 5/18}
			\node[dot] (p\i) at (\angle:1.8) {};
			
			\foreach \i/\angle in {1/90, 2/162, 3/234, 4/306, 5/18}
			\draw[probe] (0,0) -- (\angle:1.8);
			
			\node[above] at (p1) {\tiny $O_1$};
			\node[above left] at (p2) {\tiny $O_2$};
			\node[below left] at (p3) {\tiny $O_3$};
			\node[below right] at (p4) {\tiny $O_4$};
			\node[above right] at (p5) {\tiny $O_5$};
			
			\coordinate (mid51) at (54:1.8);   
			\coordinate (mid23) at (198:1.8);  
			
			\draw[arrow] (mid51) -- (mid23);
			
			
		\end{tikzpicture}
		\caption{$51\to23$}
		\label{fig:arrow-51-23}
	\end{subfigure}
	\hfill
	\begin{subfigure}[c]{0.18\textwidth}
		\centering
		\begin{tikzpicture}[
			thick,
			scale=0.45,
			dot/.style={circle, fill=black, inner sep=0.8pt},
			probe/.style={dashed, line width=0.6pt},
			arrow/.style={line width=1pt, -{Stealth[length=3pt]}, blue}
			]
			
			\draw (0,0) circle (1.8cm);
			
			\foreach \i/\angle in {1/90, 2/162, 3/234, 4/306, 5/18}
			\node[dot] (p\i) at (\angle:1.8) {};
			
			\foreach \i/\angle in {1/90, 2/162, 3/234, 4/306, 5/18}
			\draw[probe] (0,0) -- (\angle:1.8);
			
			\node[above] at (p1) {\tiny $O_1$};
			\node[above left] at (p2) {\tiny $O_2$};
			\node[below left] at (p3) {\tiny $O_3$};
			\node[below right] at (p4) {\tiny $O_4$};
			\node[above right] at (p5) {\tiny $O_5$};
			
			\coordinate (mid12) at (126:1.8);  
			\coordinate (mid34) at (270:1.8);  
			
			\draw[arrow] (mid12) -- (mid34);
			
			
		\end{tikzpicture}
		\caption{$12\to34$}
		\label{fig:arrow-12-34}
	\end{subfigure}
	\hfill
	\begin{subfigure}[c]{0.18\textwidth}
		\centering
		\begin{tikzpicture}[
			thick,
			scale=0.45,
			dot/.style={circle, fill=black, inner sep=0.8pt},
			probe/.style={dashed, line width=0.6pt},
			arrow/.style={line width=1pt, -{Stealth[length=3pt]}, blue}
			]
			
			\draw (0,0) circle (1.8cm);
			
			\foreach \i/\angle in {1/90, 2/162, 3/234, 4/306, 5/18}
			\node[dot] (p\i) at (\angle:1.8) {};
			
			\foreach \i/\angle in {1/90, 2/162, 3/234, 4/306, 5/18}
			\draw[probe] (0,0) -- (\angle:1.8);
			
			\node[above] at (p1) {\tiny $O_1$};
			\node[above left] at (p2) {\tiny $O_2$};
			\node[below left] at (p3) {\tiny $O_3$};
			\node[below right] at (p4) {\tiny $O_4$};
			\node[above right] at (p5) {\tiny $O_5$};
			
			\coordinate (mid23) at (198:1.8);  
			\coordinate (mid45) at (342:1.8);  
			
			\draw[arrow] (mid23) -- (mid45);
			
			
		\end{tikzpicture}
		\caption{$23\to45$}
		\label{fig:arrow-23-45}
	\end{subfigure}
	\hfill
	\begin{subfigure}[c]{0.18\textwidth}
		\centering
		\begin{tikzpicture}[
			thick,
			scale=0.45,
			dot/.style={circle, fill=black, inner sep=0.8pt},
			probe/.style={dashed, line width=0.6pt},
			arrow/.style={line width=1pt, -{Stealth[length=3pt]}, blue}
			]
			
			\draw (0,0) circle (1.8cm);
			
			\foreach \i/\angle in {1/90, 2/162, 3/234, 4/306, 5/18}
			\node[dot] (p\i) at (\angle:1.8) {};
			
			\foreach \i/\angle in {1/90, 2/162, 3/234, 4/306, 5/18}
			\draw[probe] (0,0) -- (\angle:1.8);
			
			\node[above] at (p1) {\tiny $O_1$};
			\node[above left] at (p2) {\tiny $O_2$};
			\node[below left] at (p3) {\tiny $O_3$};
			\node[below right] at (p4) {\tiny $O_4$};
			\node[above right] at (p5) {\tiny $O_5$};
			
			\coordinate (mid34) at (270:1.8);  
			\coordinate (mid51) at (54:1.8);   
			
			\draw[arrow] (mid34) -- (mid51);
			
			
		\end{tikzpicture}
		\caption{$34\to51$}
		\label{fig:arrow-34-51}
	\end{subfigure}
	\hfill
	\begin{subfigure}[c]{0.18\textwidth}
		\centering
		\begin{tikzpicture}[
			thick,
			scale=0.45,
			dot/.style={circle, fill=black, inner sep=0.8pt},
			probe/.style={dashed, line width=0.6pt},
			arrow/.style={line width=1pt, -{Stealth[length=3pt]}, blue}
			]
			
			\draw (0,0) circle (1.8cm);
			
			\foreach \i/\angle in {1/90, 2/162, 3/234, 4/306, 5/18}
			\node[dot] (p\i) at (\angle:1.8) {};
			
			\foreach \i/\angle in {1/90, 2/162, 3/234, 4/306, 5/18}
			\draw[probe] (0,0) -- (\angle:1.8);
			
			\node[above] at (p1) {\tiny $O_1$};
			\node[above left] at (p2) {\tiny $O_2$};
			\node[below left] at (p3) {\tiny $O_3$};
			\node[below right] at (p4) {\tiny $O_4$};
			\node[above right] at (p5) {\tiny $O_5$};
			
			\coordinate (mid45) at (342:1.8);  
			\coordinate (mid12) at (126:1.8);  
			
			\draw[arrow] (mid45) -- (mid12);
			
			
		\end{tikzpicture}
		\caption{$45\to12$}
		\label{fig:arrow-45-12}
	\end{subfigure}
	
	\caption{converting $n(\tau,\tau')$ to $g(\tau,\tau')$n the path integral five-point functions.}
	\label{fig:ten-arrow-diagrams}
\end{figure}

Using the relation \(n(\tau,\tau') = -\frac{1}{2} \partial_\tau \partial_{\tau'} g(\tau,\tau')\) and integrating by parts, we can express \(K[n]\) in terms of the bilocal field \(g_{ij} \equiv g(\tau_i,\tau_j)\):
\begin{align}
	K[n] ={}& 
	\frac{\Delta_{12} - \Delta_1 - \Delta_2}{2} \, g_{12}
	+ \frac{\Delta_{23} - \Delta_2 - \Delta_3}{2} \, g_{23}
	+ \frac{\Delta_{34} - \Delta_3 - \Delta_4}{2} \, g_{34} \nonumber \\
	&+ \frac{\Delta_{45} - \Delta_4 - \Delta_5}{2} \, g_{45}
	+ \frac{\Delta_{15} - \Delta_1 - \Delta_5}{2} \, g_{15}
	+ \frac{\Delta_2 - \Delta_{12} - \Delta_{23} + \Delta_{45}}{2} \, g_{13} \nonumber \\
	&+ \frac{\Delta_5 + \Delta_{23} - \Delta_{45} - \Delta_{15}}{2} \, g_{14}
	+ \frac{\Delta_3 - \Delta_{23} - \Delta_{34} + \Delta_{15}}{2} \, g_{24} \nonumber \\
	&+ \frac{\Delta_1 - \Delta_{12} + \Delta_{34} - \Delta_{15}}{2} \, g_{25}
	+ \frac{\Delta_4 + \Delta_{12} - \Delta_{34} - \Delta_{45}}{2} \, g_{35}.
	\label{eq:K-in-terms-of-g}
\end{align}
Figure \ref{fig:ten-arrow-diagrams} Schematic diagrams for converting $n(\tau,\tau')$ to $g(\tau,\tau')$ in the five-point path integral. \\

\textbf{Result in the zero-temperature limit}

Evaluating the path integral in the saddle-point approximation and taking the zero-temperature limit where \(g_{ij} = -\log |\tau_{ij}|^2\), we obtain the explicit five-point contact correlator:
\begin{align}
	\langle O_5 O_4 O_3 O_2 O_1 \rangle_{\text{contact}} 
	={}& \frac{1}{N} 
	|\tau_{12}|^{\Delta_{12} - \Delta_1 - \Delta_2}
	|\tau_{23}|^{\Delta_{23} - \Delta_2 - \Delta_3}
	|\tau_{34}|^{\Delta_{34} - \Delta_3 - \Delta_4}
	|\tau_{45}|^{\Delta_{45} - \Delta_4 - \Delta_5} \nonumber \\
	&\times 
	|\tau_{15}|^{\Delta_{15} - \Delta_1 - \Delta_5} 
	|\tau_{13}|^{\Delta_{45} - \Delta_{23} - \Delta_{12} + \Delta_2}
	|\tau_{14}|^{\Delta_{23} - \Delta_{45} - \Delta_{15} + \Delta_5}  \nonumber \\
	&\times 
	|\tau_{24}|^{\Delta_{15} - \Delta_{23} - \Delta_{34} + \Delta_3} 
	|\tau_{25}|^{\Delta_{34} - \Delta_{15} - \Delta_{12} + \Delta_1}
	|\tau_{35}|^{\Delta_{12} - \Delta_{34} - \Delta_{45} + \Delta_4}.
	\label{eq:5pt-explicit}
\end{align}
The exponents $\gamma_{ij}$ are listed in Table~\ref{tab:5pt_exponents}.

To exhibit the dependence on the two independent cross-ratios, we set
\begin{equation}
	\tau_1 = 0,\quad \tau_2 = x,\quad \tau_3 = y,\quad \tau_4 = 1,\quad \tau_5 \to \infty,
\end{equation}
and define the scaled correlator
\begin{equation}
	I(x,y) \coloneqq \lim_{\tau_5 \to \infty} |\tau_5|^{2\Delta_5} \,
	\langle O_5 O_4 O_3 O_2 O_1 \rangle_{\text{contact}} .
\end{equation}
Then
\begin{align}
	I(x,y) ={}& \frac{1}{N} \,
	|x|^{\Delta_{12} - \Delta_1 - \Delta_2}
	\, |y-x|^{\Delta_{23} - \Delta_2 - \Delta_3}
	\, |1-y|^{\Delta_{34} - \Delta_3 - \Delta_4} \nonumber \\
	& \times |y|^{\Delta_{45} - \Delta_4 - \Delta_5}
	\, |1-x|^{\Delta_{15} - \Delta_1 - \Delta_5}.
	\label{eq:5pt-cross-ratios}
\end{align}
This expression manifests the expected factorized power-law form dictated by conformal symmetry for a five-point contact diagram. The exponents are linear combinations of the flux parameters \(\Delta_i\) and \(\Delta_{ij}\), which can be tuned to match specific AdS$_2$ scalar contact Witten diagrams.

\noindent
For convenience, we list the exponents $\gamma_{ij} = -2\, \text{coefficient of } \ln|\tau_{ij}|$ in the five-point correlator $\langle O_5\cdots O_1\rangle_{\text{contact}} \propto \prod_{i<j} |\tau_{ij}|^{-\gamma_{ij}}$:

\begin{table}[h]
	\centering
	\begin{tabular}{c|c}
		$ij$ & $\gamma_{ij}$ \\ \hline
		12 & $\Delta_{12} - \Delta_1 - \Delta_2$ \\
		23 & $\Delta_{23} - \Delta_2 - \Delta_3$ \\
		34 & $\Delta_{34} - \Delta_3 - \Delta_4$ \\
		45 & $\Delta_{45} - \Delta_4 - \Delta_5$ \\
		15 & $\Delta_{15} - \Delta_1 - \Delta_5$ \\
		13 & $\Delta_{45} - \Delta_{23} - \Delta_{12} + \Delta_2$ \\
		14 & $\Delta_{23} - \Delta_{45} - \Delta_{15} + \Delta_5$ \\
		24 & $\Delta_{15} - \Delta_{23} - \Delta_{34} + \Delta_3$ \\
		25 & $\Delta_{34} - \Delta_{15} - \Delta_{12} + \Delta_1$ \\
		35 & $\Delta_{12} - \Delta_{34} - \Delta_{45} + \Delta_4$
	\end{tabular}
	\caption{Exponents $\gamma_{ij}$ for the five-point contact correlator.}
	\label{tab:5pt_exponents}
\end{table}

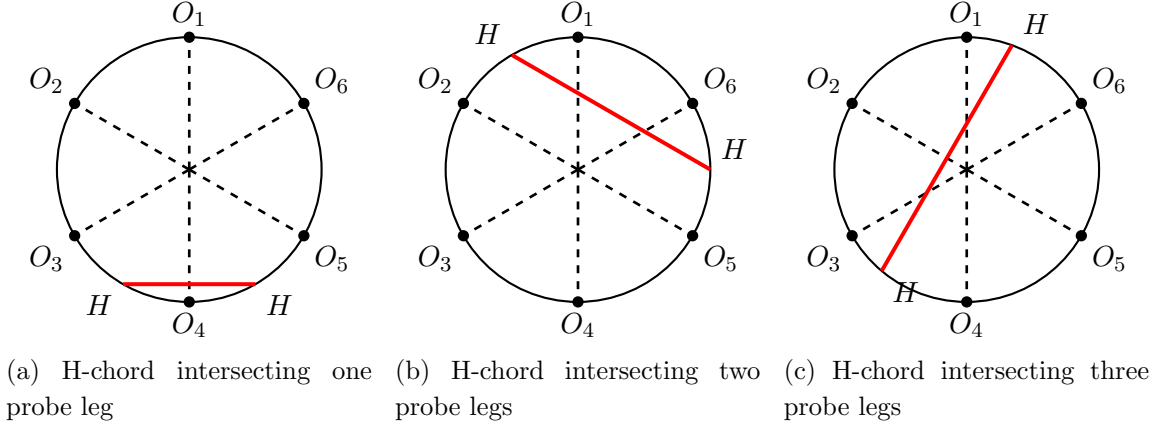
\begin{figure}[htbp]
	\centering
	
	\begin{subfigure}[c]{0.32\textwidth}
		\centering
		\begin{tikzpicture}[
			thick,
			scale=0.7,
			dot/.style={circle, fill=black, inner sep=1.5pt},
			probe/.style={dashed, line width=1pt},
			Hchord/.style={line width=1.5pt, red}
			]
			
			\useasboundingbox (-3.2,-3.2) rectangle (3.2,3.2);
			
			\draw (0,0) circle (2.5cm);
			
			\foreach \i/\angle in {1/90, 2/150, 3/210, 4/270, 5/330, 6/30}
			\node[dot] (p\i) at (\angle:2.5) {};
			
			\foreach \i/\angle in {1/90, 2/150, 3/210, 4/270, 5/330, 6/30}
			\draw[probe] (0,0) -- (\angle:2.5);
			
			\node[above] at (p1) {$O_1$};
			\node[above left] at (p2) {$O_2$};
			\node[below left] at (p3) {$O_3$};
			\node[below] at (p4) {$O_4$};
			\node[below right] at (p5) {$O_5$};
			\node[above right] at (p6) {$O_6$};
			
			\coordinate (Hstart) at (240:2.5);    
			\coordinate (Hend) at (300:2.5);      
			
			\draw[Hchord] (Hstart) -- (Hend);
			
			\node[below left] at (Hstart) {$H$};
			\node[below right] at (Hend) {$H$};

		\end{tikzpicture}
		\caption{H-chord intersecting one probe leg}
		\label{fig:six-point-single}
	\end{subfigure}
	\hfill
	\begin{subfigure}[c]{0.32\textwidth}
		\centering
		\begin{tikzpicture}[
			thick,
			scale=0.7,
			dot/.style={circle, fill=black, inner sep=1.5pt},
			probe/.style={dashed, line width=1pt},
			Hchord/.style={line width=1.5pt, red}
			]
			
			\useasboundingbox (-3.2,-3.2) rectangle (3.2,3.2);
			
			\draw (0,0) circle (2.5cm);
			
			\foreach \i/\angle in {1/90, 2/150, 3/210, 4/270, 5/330, 6/30}
			\node[dot] (p\i) at (\angle:2.5) {};
			
			\foreach \i/\angle in {1/90, 2/150, 3/210, 4/270, 5/330, 6/30}
			\draw[probe] (0,0) -- (\angle:2.5);
			
			\node[above] at (p1) {$O_1$};
			\node[above left] at (p2) {$O_2$};
			\node[below left] at (p3) {$O_3$};
			\node[below] at (p4) {$O_4$};
			\node[below right] at (p5) {$O_5$};
			\node[above right] at (p6) {$O_6$};
			
			\coordinate (Hstart) at (120:2.5);    
			\coordinate (Hend) at (0:2.5);        
			
			\draw[Hchord] (Hstart) -- (Hend);
			
			\node[above left] at (Hstart) {$H$};
			\node[above right] at (Hend) {$H$};

		\end{tikzpicture}
		\caption{H-chord intersecting two probe legs}
		\label{fig:six-point-double}
	\end{subfigure}
	\hfill
	\begin{subfigure}[c]{0.32\textwidth}
		\centering
		\begin{tikzpicture}[
			thick,
			scale=0.7,
			dot/.style={circle, fill=black, inner sep=1.5pt},
			probe/.style={dashed, line width=1pt},
			Hchord/.style={line width=1.5pt, red}
			]
			
			\useasboundingbox (-3.2,-3.2) rectangle (3.2,3.2);
			
			\draw (0,0) circle (2.5cm);
			
			\foreach \i/\angle in {1/90, 2/150, 3/210, 4/270, 5/330, 6/30}
			\node[dot] (p\i) at (\angle:2.5) {};
			
			\foreach \i/\angle in {1/90, 2/150, 3/210, 4/270, 5/330, 6/30}
			\draw[probe] (0,0) -- (\angle:2.5);
			
			\node[above] at (p1) {$O_1$};
			\node[above left] at (p2) {$O_2$};
			\node[below left] at (p3) {$O_3$};
			\node[below] at (p4) {$O_4$};
			\node[below right] at (p5) {$O_5$};
			\node[above right] at (p6) {$O_6$};
			
			\coordinate (Hstart) at (70:2.5);     
			\coordinate (Hend) at (230:2.5);      
			
			\draw[Hchord] (Hstart) -- (Hend);
			
			\node[above right] at (Hstart) {$H$};
			\node[below right] at (Hend) {$H$};

		\end{tikzpicture}
		\caption{H-chord intersecting three probe legs}
		\label{fig:six-point-triple}
	\end{subfigure}
	
	\caption{Six-point contact chord diagrams. (a) A  H-chord intersecting probe $O_4$. (b) A  H-chord intersecting probes $O_1$ and $O_6$. (c) A H-chord intersecting probes $O_1$, $O_2$ and $O_3$. Dashed lines represent probe operators $O_i$, red lines represent H-chords.}
	\label{fig:six-point-contact}
\end{figure}

For six points the set of configurations grows further, with $L=2$ admitting all sequences with an even number of $+$ signs, $L=3$ those with sum divisible by 3, etc. In both cases, larger $L$ provides a richer set of sign patterns, which in turn increases the variety of bulk contact vertices we can represent.

\par The chord path integral computation for five and six points becomes substantially more involved because the number of independent cross‑ratios grows and the chord diagrams can contain more complicated intersection patterns. In particular, for six‑point functions one must account for triple intersections of an $H$‑chord with three different probe chords, which give rise to factors of the form
\begin{equation}
	q^{\Delta_{ijk}} = \Big\langle \cos\!\Big( \frac{3F + \tilde{F}^{(i)} + \tilde{F}^{(j)} + \tilde{F}^{(k)}}{2} \Big) \Big\rangle \quad (\epsilon_i = \epsilon_j = \epsilon_k),
\end{equation}
or similar expressions when the signs are not all equal. These triple‑joint parameters $\Delta_{ijk}$ enter the exponents $\gamma_{ij}^{(\mathcal{C})}$ and enlarge the parameter space available for matching.

\par For a fixed configuration $\mathcal{C}$, the chord path integral again yields a purely factorized power‑law structure (in the zero‑temperature limit):
\begin{align}
    \langle O_5 \cdots O_1 \rangle_{\mathcal{C}} &\sim \frac{1}{N^{3/2}} \prod_{1\le i<j\le 5} \frac{1}{|\tau_{ij}|^{\gamma_{ij}^{(\mathcal{C})}}},\nonumber \\
    \langle O_6 \cdots O_1 \rangle_{\mathcal{C}} &\sim \frac{1}{N^{2}} \prod_{1\le i<j\le 6} \frac{1}{|\tau_{ij}|^{\gamma_{ij}^{(\mathcal{C})}}},
\end{align}
where the exponents $\gamma_{ij}^{(\mathcal{C})}$ are linear combinations of the flux parameters $\Delta_i^{(\mathcal{C})}$, $\Delta_{ij}^{(\mathcal{C})}$ and, for six points, also of triple‑joint factors $\Delta_{ijk}^{(\mathcal{C})}$ that arise when an $H$‑chord intersects three different probe chords simultaneously.

\section{Linear Combinations of Configurations}
\label{sec:linear_comb}

\par A major limitation of the original chord diagram constructions is that they only produce pure power-law functions 
\begin{eqnarray}
    x^a (1-x)^b.
\end{eqnarray}
However, many AdS$_2$ contact Witten diagrams~\cite{Bliard:2022xsm} contain logarithmic singularities, e.g., terms like $x^a \log x$ or $(1-x)^b \log(1-x)$. To capture such terms, we introduce the technique of parameter degeneracy.

Consider two configurations $\mathcal{C}_1$ and $\mathcal{C}_2$ that respectively yield exponents $(a_1, b)$ and $(a_2, b)$ with $a_1 \neq a_2$. Their combined contribution (with weights $w_1, w_2$) is
\begin{equation}
	w_1 x^{a_1} (1-x)^b + w_2 x^{a_2} (1-x)^b.
\end{equation}
Now suppose we can adjust the microscopic parameters (flux distributions) such that, 
\begin{eqnarray}
    a_2 = a_1 + \epsilon
\end{eqnarray}
with $\epsilon$ small. Moreover, let the weights be tuned to be 
\begin{eqnarray}
    w_1 = A/\epsilon, \quad w_2 = -A/\epsilon.
\end{eqnarray}
Then the sum becomes
\begin{equation}
	\frac{A}{\epsilon} x^{a_1} (1-x)^b (x^\epsilon - 1) = A x^{a_1} (1-x)^b \frac{x^\epsilon - 1}{\epsilon}.
\end{equation}
Taking the limit $\epsilon \to 0$ and using $\lim_{\epsilon\to0} (x^\epsilon - 1)/\epsilon = \log x$, we obtain
\begin{equation}
	A x^{a_1} (1-x)^b \log x.
\end{equation}
Thus, by carefully taking a limit where the exponents of two configurations become degenerate and the weights diverge appropriately, we can generate logarithmic terms. Similarly, degeneracy in the $b$ exponent produces $\log(1-x)$.

Physically, such a degeneracy limit may correspond to a fine-tuned point in the space of bulk couplings where two different contact vertices contribute with equal amplitude, leading to enhanced quantum interference. The logarithmic singularities are reminiscent of those that appear at critical points in statistical mechanics.

\par In a generic physical situation, the bulk contact interaction may involve a combination of different derivative couplings. Therefore, the boundary correlation function should receive contributions from all possible contact vertices compatible with the symmetries. In our framework, this translates to taking a \emph{linear combination} of the contributions from all allowed probe sign configurations:
\begin{equation}
	I_{\text{phys}}(x) = \sum_{\mathcal{C}} w_{\mathcal{C}} \, I_{\mathcal{C}}(x) = \frac{1}{N} \sum_{\mathcal{C}} w_{\mathcal{C}} \, x^{a_{\mathcal{C}}} (1-x)^{b_{\mathcal{C}}},
\end{equation}
where the weights $w_{\mathcal{C}}$ are determined by the microscopic details of the model (such as the specific flux distributions and the relative probabilities of different configurations). In the absence of further information, we treat $w_{\mathcal{C}}$ as free parameters, subject to the normalization $\sum_{\mathcal{C}} w_{\mathcal{C}} = 1$.

The crucial observation is that even for a fixed set of conformal dimensions $\Delta_i$, the exponents $a_{\mathcal{C}}$ and $b_{\mathcal{C}}$ can vary between configurations because $\Delta_{12}^{(\mathcal{C})}$ and $\Delta_{23}^{(\mathcal{C})}$ depend on the signs of the probes. Thus, the linear combination spans a multi-dimensional space of functions, much richer than the single power-law function obtained from any single configuration.

As an illustrative example, consider the Witten diagram $I_{[2,2,1,1]}(x)$, which has the form~\cite{Bliard:2022xsm}
\begin{equation}
	I_{[2,2,1,1]}(x) = c \left[ \frac{\log(1-x)}{x} + \frac{\log x}{1-x} \right].
\end{equation}
To reproduce it, we work on an $L=3$ lattice and use the two configurations $\mathcal{C}_1 = (+,+,-,-)$ and $\mathcal{C}_2 = (+,-,-,+)$. By choosing flux parameters such that$\Delta_1 = \Delta_2 = 2, \quad \Delta_3 = \Delta_4 = 1,\Delta_{12}^{(\mathcal{C}_1)} = 3, \Delta_{23}^{(\mathcal{C}_1)} = 0,  \Delta_{12}^{(\mathcal{C}_2)} = 3, \Delta_{23}^{(\mathcal{C}_2)} = \epsilon$
we obtain $a_{\mathcal{C}_1}=a_{\mathcal{C}_2}=-1$, $b_{\mathcal{C}_1}=0$, $b_{\mathcal{C}_2}=\epsilon$. Setting $w_{\mathcal{C}_1} = A/\epsilon$ and $w_{\mathcal{C}_2} = -A/\epsilon$, the limit $\epsilon \to 0$ yields
\begin{equation}
	I_{\text{phys}}(x) \longrightarrow \frac{A}{N} \, x^{-1} \log(1-x).
\end{equation}
A similar construction with another pair of configurations (or by exploiting symmetry $x \leftrightarrow 1-x$) generates the term $x^{-1} \log x$, thus matching the desired Witten diagram up to an overall constant.

We now summarize the outcome of our systematic matching efforts for the scalar contact Witten diagrams listed in~\cite{Bliard:2022xsm}. The key insight is that, by taking linear combinations of contributions from different configurations and tuning the flux parameters to degeneracy points, we can generate not only pure power laws but also terms with logarithmic singularities and rational functions. In principle, given a sufficiently large set of independent configurations and provided the positivity constraints are satisfied, we can match any contact Witten diagram through an appropriate linear combination.

The reason is that the space of functions spanned by linear combinations of the form  is dense in the class of functions that appear in AdS$_2$ contact Witten diagrams. For four‑point functions, the available configurations are limited but already numerous enough to produce a wide variety of functional forms. For example:
\begin{itemize}
	\item Pure power‑law diagrams such as $I_{[1,1,1,2]}(x) \propto x^{-1}$ and $I_{[2,2,2,1]}(x) \propto [x(1-x)]^{-1}$ are matched exactly by a single configuration (typically the alternating one $(+,-,+,-)$ on an $L=2$ lattice) with suitably chosen $\Delta_i$ and $\Delta_{ij}$.
	\item Diagrams containing logarithmic singularities, such as $I_{[2,2,1,1]}(x)$, can be obtained by taking a linear combination of two configurations whose exponents are tuned to a degeneracy point, as demonstrated above.
	\item More intricate diagrams that are rational functions of $x$ (sums of several power laws) can be reproduced by combining several configurations with different exponents, without requiring a degeneracy limit.
\end{itemize}

Crucially, every match we have presented respects the positivity constraints derived from the covariance matrix of the flux variables. This ensures that the corresponding flux distributions are physically realizable. For instance, in the logarithmic example above, One can verify that for sufficiently small $\epsilon$ the $4\times4$ covariance matrix of $\{a_i\}$ can be chosen to remain positive semi-definite. Hence the required fine-tuning is realizable within our Gaussian framework.

Therefore, our generalized chord diagram technique, combined with the linear‑combination and parameter‑degeneracy approach, provides a systematic and physically consistent way to reproduce all known scalar contact Witten diagrams in AdS$_2$. The method becomes even more powerful for higher‑point functions, where the number of independent configurations grows with the lattice size $L$, further expanding the function space we can access.

The linear combination of configurations and the parameter degeneracy limit can be incorporated into the chord path integral formalism. The partition function becomes a sum over configurations $\mathcal{C}$, each weighted by $w_{\mathcal{C}}$, and the action for each configuration depends on the parameters $\Delta_i^{(\mathcal{C})}$, $\Delta_{ij}^{(\mathcal{C})}$, etc. In the degeneracy limit, the path integral will involve nearly canceling contributions whose interplay produces logarithms.

We consider another example of a five-point function, $I_{[2,1,1,2,1]}$~\cite{Bliard:2022xsm}:
\begin{equation}
	I_{[2,1,1,2,1]}=-\frac{\pi^{2}(2 x y+x-3 y)}{16(x-1)^{2} y^{2}}.
	\label{exp_5pt}
\end{equation}

To illustrate this matching explicitly, we decompose the rational function in Eq.~(5.7) into a sum of three pure power-law terms multiplied by an overall factor:
\begin{equation}
    I_{[2,1,1,2,1]} = -\frac{\pi^2}{16} \Bigl[ 2\, x \, y^{-1} (1-x)^{-2} + x \, y^{-2} (1-x)^{-2} - 3\, y^{-1} (1-x)^{-2} \Bigr].
    \label{eq:decompose5}
\end{equation}
We will match each term with a specific chord diagram configuration on a lattice of size $L=3$, for which the closure condition $\sum_{i=1}^5 \epsilon_i \equiv 0 \pmod 3$ admits the solutions with four $+1$ and one $-1$ (or the overall sign-flipped counterparts). 
The three independent configurations, labeled by the position $k$ of the single minus sign, are
\begin{equation}
    \mathcal{C}_1:\; (+,+,+,+,-)\;(k=5),\qquad
    \mathcal{C}_2:\; (+,+,+,-,+)\;(k=4),\qquad
    \mathcal{C}_3:\; (+,+,-,+,+)\;(k=3).
    \label{eq:five_configs}
\end{equation}
Each sequence sums to $3\equiv0\pmod 3$, thus satisfying the closure condition.

From the Witten diagram label $[2,1,1,2,1]$, the conformal dimensions of the five probe operators are fixed to be
\begin{equation}
    \Delta_1=2,\quad \Delta_2=1,\quad \Delta_3=1,\quad \Delta_4=2,\quad \Delta_5=1.
    \label{eq:deltas_2121}
\end{equation}
To eliminate the dependence on the cross ratios $|y-x|$ and $|1-y|$, we impose for all configurations
\begin{equation}
    \Delta_{23} = \Delta_2 + \Delta_3 = 2,\qquad \Delta_{34} = \Delta_3 + \Delta_4 = 3.
    \label{eq:zero_cond}
\end{equation}
The remaining independent two-index parameters $\Delta_{12},\Delta_{45},\Delta_{15}$ are then chosen to reproduce the three monomials in Eq.~\eqref{eq:decompose5}. The required values are listed in Table~\ref{tab:params5}.

\begin{table}[h]
\centering
\caption{Parameter choices for the three configurations matching $I_{[2,1,1,2,1]}$. The single-probe dimensions are fixed by Eq.~\eqref{eq:deltas_2121}, and $\Delta_{23}=2$, $\Delta_{34}=3$ by Eq.~\eqref{eq:zero_cond}.}
\label{tab:params5}
\begin{tabular}{c|c|c|c}
Configuration & $\Delta_{12}$ & $\Delta_{45}$ & $\Delta_{15}$ \\ \hline
$\mathcal{C}_1\;(k=5)$ & 4 & 2 & 1 \\
$\mathcal{C}_2\;(k=4)$ & 4 & 1 & 1 \\
$\mathcal{C}_3\;(k=3)$ & 3 & 2 & 1
\end{tabular}
\end{table}

With these parameters, the correlators evaluate to
\begin{align}
    I_{\mathcal{C}_1} &= \frac{1}{N} \, x^{4-2-1} \, y^{2-2-1} \, (1-x)^{1-2-1} = \frac{1}{N} \, x^1 y^{-1} (1-x)^{-2}, \nonumber\\
    I_{\mathcal{C}_2} &= \frac{1}{N} \, x^{1} y^{1-2-1} (1-x)^{-2} = \frac{1}{N} \, x^1 y^{-2} (1-x)^{-2}, \\
    I_{\mathcal{C}_3} &= \frac{1}{N} \, x^{3-2-1} y^{2-2-1} (1-x)^{-2} = \frac{1}{N} \, x^0 y^{-1} (1-x)^{-2}. \nonumber
\end{align}

We now verify that these parameter sets satisfy the positivity constraints. 
The flux parameters are encoded in the centered variables $a_i = (F+\tilde F^{(i)})/(2\sqrt{\langle F^2\rangle})$, $i=1,\dots,5$, and the $\Delta_{ij}$ are related to the inner products $\langle a_i a_j\rangle$ through
\begin{equation}
    \Delta_{ij} = 
    \begin{cases}
        \Delta_i + \Delta_j - 2\langle a_i a_j\rangle, & \epsilon_i = \epsilon_j,\\[2pt]
        \Delta_i + \Delta_j + 2\langle a_i a_j\rangle, & \epsilon_i \neq \epsilon_j.
    \end{cases}
    \label{eq:inner_prod}
\end{equation}
The physical realizability of the flux parameters is equivalent to the existence of real vectors $\mathbf{a}_i$ in some Euclidean space whose inner products reproduce these $\langle a_i a_j\rangle$, i.e.\ the Gram matrix $G_{ij}=\langle a_i a_j\rangle$ must be positive semi-definite.
For each configuration, we can explicitly construct such vectors in $\mathbb{R}^d$ ($d\le5$), confirming that all leading principal minors of $G$ are positive.

For instance, for $\mathcal{C}_1$ one can choose
\begin{align}
    \mathbf{a}_1 &= (\sqrt{2},\,0,\,0,\,0), \nonumber\\
    \mathbf{a}_2 &\approx (-0.3535,\,0.9354,\,0,\,0), \nonumber\\
    \mathbf{a}_3 &= (0,\,0,\,1,\,0), \nonumber\\
    \mathbf{a}_4 &= (0,\,0,\,0,\,\sqrt{2}), \nonumber\\
    \mathbf{a}_5 &\approx (-0.7071,\,0.3535,\,0.4082,\,-0.3535),
    \label{eq:vectors}
\end{align}
which satisfies $\langle \mathbf{a}_1,\mathbf{a}_2\rangle = -0.5$, $\langle \mathbf{a}_4,\mathbf{a}_5\rangle = -0.5$, $\langle \mathbf{a}_1,\mathbf{a}_5\rangle = -1$, and $\|\mathbf{a}_i\|^2 = \Delta_i$. The Gram matrix is positive definite, confirming the physical realizability of the flux distribution. The same construction works for $\mathcal{C}_2$ and $\mathcal{C}_3$ with appropriate vectors.

The physical correlator is the weighted sum $I_{\rm phys} = \sum_{\mathcal{C}} w_{\mathcal{C}} I_{\mathcal{C}}$. Using Eq.~\eqref{eq:decompose5}, we read off the weights
\begin{equation}
    w_{\mathcal{C}_1} = -2\cdot\frac{\pi^2}{16} N,\qquad
    w_{\mathcal{C}_2} = -\frac{\pi^2}{16} N,\qquad
    w_{\mathcal{C}_3} = 3\cdot\frac{\pi^2}{16} N.
\end{equation}
Summing the three contributions precisely reproduces $I_{[2,1,1,2,1]}$:
\begin{equation}
    I_{\rm phys}(x,y) = w_{\mathcal{C}_1} I_{\mathcal{C}_1} + w_{\mathcal{C}_2} I_{\mathcal{C}_2} + w_{\mathcal{C}_3} I_{\mathcal{C}_3} = -\frac{\pi^2(2xy + x - 3y)}{16(x - 1)^2 y^2}.
\end{equation}
This demonstrates that even complicated rational contact Witten diagrams can be systematically assembled from a small number of chord configurations, each contributing a simple power-law factor.

It is important to stress the limitations of the current framework.  
Because we restrict ourselves to Gaussian flux distributions, the exponents $\Delta_i$ and $\Delta_{ij}$ are quadratic in the fluxes and the resulting correlators are always (finite sums of) power-laws times possible logarithmic factors.  
Contact Witten diagrams that involve higher-order transcendental functions, such as polylogarithms or generalized hypergeometric functions, cannot be obtained from Gaussian fluxes alone; their matching would require non-Gaussian statistics with higher-order cumulants.  
Furthermore, the degeneracy mechanism only produces logarithms of the cross ratios; functions like $(\log x)^2$ or ${\rm Li}_2(x)$ are not accessible with the current setup.  
In addition, the closure condition $\sum_i\epsilon_i\equiv0\pmod L$ restricts the possible exponent combinations; for a given $L$, only a finite-dimensional subspace of the full function space is spanned.  
A systematic classification of which Witten diagrams can (or cannot) be matched within this approach, perhaps in terms of the complexity of the required flux distributions, would be a useful extension.  
We leave such a study for future work.

The above matching procedure illustrates the general strategy. 
The lattice size $L$ controls the set of allowed probe sign configurations via the closure condition $\sum_i \epsilon_i \equiv 0 \pmod L$. While the number of configurations does not grow monotonically with $L$ for fixed $n$, increasing $L$ expands the range of accessible sign patterns across different $n$, thereby enriching the functional basis for the linear combination scheme. 
In the $L\to\infty$ limit, the sum over configurations approximates an integral over a continuous family of boundary exponents, making the function space spanned by chord diagrams dense in the class of AdS$_2$ contact Witten diagrams.

The parameter degeneracy mechanism, which generates logarithmic singularities via limits of the form $\epsilon^{-1}(x^\epsilon-1)\to\log x$, admits a direct bulk interpretation: the degeneracy of conformal exponents corresponds to the confluence of operator dimensions, while the diverging weights $w_{\mathcal{C}}\sim 1/\epsilon$ reflect the emergence of logarithmic behavior at marginal couplings. It is suggestive that this fine-tuning in the space of flux distributions parallels the renormalization group flows that generate logarithmic terms in the bulk effective action.

From a microscopic perspective, the weights $w_{\mathcal C}$ arise naturally when the flux distributions for different probe operators are not fully fixed but instead appear with certain probabilities in an ensemble. Equivalently, the effective action of the boundary theory may contain several independent contact vertices, each associated with a specific sign configuration $\mathcal C$ and a coupling constant $g_{\mathcal C}$. In the large-$N$ limit, the correlation function is a linear combination of contributions from each vertex, weighted by $g_{\mathcal C}$. In this work we treat $w_{\mathcal C}$ as phenomenological coefficients; determining them from first principles would require a more detailed specification of the flux distributions, which we leave for future investigation.

\section{Discussion and Conclusions}
\label{sec:discussion}

In this work we have developed a systematic framework for constructing contact chord diagrams in a broader class of Fock-space flux models, moving beyond earlier constructions that were 
limited to specific probe sign configurations~\cite{Berkooz:2023cqc,Berkooz:2023scv}. 
By introducing an arbitrary periodic lattice size $L$ and formulating the physical correlation function as a weighted linear combination over all allowed configurations, we have shown that chord diagrams can reproduce a comprehensive library of AdS$_2$ scalar contact Witten diagrams~\cite{Bliard:2022xsm}, including those with logarithmic singularities and rational function dependencies previously inaccessible to chord techniques.

The central element of our construction is the direct correspondence between the lattice size $L$, the probe sign configurations satisfying $\sum_i \epsilon_i = 0 \pmod L$, and the variety of accessible bulk contact vertices. We have demonstrated through explicit three- to six-point computations how linear combinations 
of the basis elements provided by different configurations, augmented by the parameter degeneracy mechanism $\lim_{\epsilon\to0} \epsilon^{-1}(x^\epsilon-1)=\log x$, span the full function space of scalar contact Witten diagrams. The successful matching of the five-point rational function $I_{[2,1,1,2,1]}$ and the outline for six-point examples confirm that the method extends to higher multiplicities without conceptual obstruction. 
All parameter choices respect the positivity constraints derived from the underlying flux covariance matrix, guaranteeing physical realizability.

The lattice size $L$ plays a role analogous to a UV regulator in the boundary theory. In the microscopic model, $L$ determines the number of sites in each Fock-space direction and, through the closure condition $\sum_i \epsilon_i = 0 \pmod L$, controls the set of allowed probe sign configurations. {While the number of allowed configurations does not grow monotonically with $L$ for fixed $n$---for example, the all-plus five-point configuration $(+,+,+,+,+)$ is allowed only at $L=5$, and for $n=4$ the set of configurations saturates at $L=4$---increasing $L$ nevertheless expands the range of possible sign patterns across different $n$, thereby providing a richer functional basis for the linear combination scheme.}
In the $L\to\infty$ limit, the sum over configurations approximates an integral over a continuous family of boundary exponents, which is precisely what one expects from a fully local bulk field theory where contact vertices can have continuously varying derivative couplings. Concretely, the spacing between allowed exponents sets an effective resolution scale; when $L$ is finite, only a discrete set of bulk derivative structures can be resolved. This suggests that $L$ should be identified with some UV scale in the bulk, perhaps related to the number of species or the volume of the internal manifold in a compactification. Verifying this conjecture would require a precise dictionary between the flux parameters and the bulk Planck scale, which we leave for future work.

These results indicate that chord diagrams may serve as a complete microscopic language for bulk contact interactions in NAdS$_2$/CFT$_1$. The required linear combinations translate directly into a sum over distinct derivative couplings in the bulk effective action, with the flux parameters $\Delta_i$ and $\Delta_{ij}$ playing the role of coupling constants. The parameter degeneracy limits that generate logarithmic terms correspond to the confluence of operator dimensions and the appearance of marginal couplings, providing a microscopic origin for the logarithmic structures often encountered in Witten diagrams.

Several limitations of the present analysis point to interesting directions for future work. {First, the entire construction relies on the assumption of Gaussian flux distributions, as introduced in Section~\ref{sec:model}; extending the framework beyond this non-trivial condition could unlock more exotic transcendental functions, such as polylogarithms, that appear in higher-loop bulk diagrams.} Second, incorporating exchange diagrams---corresponding to chord diagrams with internal splittings and joinings---is necessary for reconstructing the full bulk dynamics, not simply contact vertices. Third, it has been argued that SYK is dual to Jackiw-Teitelboim (JT) gravity \cite{Teitelboim:1983ux,Jackiw:1984je} according to AdS/CFT correspondence \cite{Almheiri:2014cka,Maldacena:2016upp,Chowdhury:2021qpy}.
a more detailed comparison with the low-energy effective action of JT gravity coupled to matter would solidify the bulk interpretation of the configuration weights $w_{\mathcal{C}}$ and the lattice size $L$, which we have conjectured may serve as a UV regulator whose large-$L$ limit restores the full local field theory in AdS$_2$. Finally, the conceptual structure of random fluxes on a discrete space generating an emergent geometry is suggestive of generalizations beyond two dimensions, perhaps through higher-rank tensor models or multi-index generalizations of the Fock-space flux model.

To conclude, we have transformed chord diagrams from a model-specific computational tool into a more general framework for cataloguing and reconstructing bulk locality in the simplest holographic setting. The explicit matching of three- to six-point contact correlators, the introduction of the linear combination and parameter degeneracy techniques, and the established link between lattice size and bulk vertex variety provide a principled roadmap for further explorations of the microscopic origins of spacetime.

\appendix
\section{Flux quantization under periodic boundary conditions}

\textbf{Representation and algebra}

Consider an $L$-dimensional representation of the $su(2)$ algebra.  
Choose the standard basis $\{|s_m\rangle\}_{m=1}^{L}$ where $\mathbb{S}^3$ is diagonal with eigenvalues
\[
s_m = \frac{L+1}{2}-m,\qquad m=1,2,\dots,L,
\]
i.e.
\[
\mathbb{S}^3 = \operatorname{diag}\left( \frac{L-1}{2},\; \frac{L-3}{2},\; \dots,\; -\frac{L-1}{2} \right).
\]
Define the raising and lowering operators $\mathbb{S}^+$ and $\mathbb{S}^-=(\mathbb{S}^+)^\dagger$ by
\[
(\mathbb{S}^+)_{m,m+1}=1,\quad m=1,\dots,L-1,\qquad \text{all other entries zero}.
\]
Thus $\mathbb{S}^+$ is the matrix with ones on the first superdiagonal.  
This normalization (dropping the square‑root factors of standard $su(2)$) is chosen to make the periodicity operator simple; it does not affect the flux quantization derived below.

One readily checks the commutation relations:
\[
[\mathbb{S}^3,\mathbb{S}^+]=\mathbb{S}^+,\qquad [\mathbb{S}^3,\mathbb{S}^-]=-\mathbb{S}^-.
\]
Indeed, the only non‑zero matrix elements of $[\mathbb{S}^3,\mathbb{S}^+]$ are
\[
(\mathbb{S}^3\mathbb{S}^+-\mathbb{S}^+\mathbb{S}^3)_{m,m+1}=s_m\cdot1-1\cdot s_{m+1}=s_m-s_{m+1}=1,
\]
hence $[\mathbb{S}^3,\mathbb{S}^+]=\mathbb{S}^+$.  The proof for $\mathbb{S}^-$ is analogous. \\

\textbf{Periodicity operator}

Define the periodicity operator
\[
P^+ = \mathbb{S}^+ + (\mathbb{S}^-)^{L-1}.
\]
Because $(\mathbb{S}^-)^{L-1}$ has a single non‑zero entry $(\mathbb{S}^-)^{L-1}_{1,L}=1$ (it maps $|s_L\rangle$ to $|s_1\rangle$), $P^+$ becomes the $L\times L$ cyclic permutation matrix
\[
P^+ = \begin{pmatrix}
	0 & 1 & 0 & \cdots & 0 \\
	0 & 0 & 1 & \cdots & 0 \\
	\vdots & \vdots & \ddots & \ddots & \vdots \\
	0 & 0 & \cdots & 0 & 1 \\
	1 & 0 & \cdots & 0 & 0
\end{pmatrix},
\]
which satisfies $(P^+)^L = \mathbb{I}_L$, i.e. periodic boundary conditions. \\

\textbf{Flux quantization}

To construct fluxed operators
\[
T_\mu^\pm = P_\mu^\pm \prod_{\rho\neq\mu} e^{\pm i\frac{F_{\mu\rho}}{2}\mathbb{S}_\rho^3}.
\]
that obey the magnetic algebra $T_\mu^+T_\nu^+ = e^{iF_{\mu\nu}} T_\nu^+T_\mu^+$, one needs the compatibility condition
\[
e^{i\frac{F}{2}\mathbb{S}^3} P^+ e^{-i\frac{F}{2}\mathbb{S}^3} = e^{i\frac{F}{2}} P^+. \tag{A.1}
\]
Evaluating the left‑hand side matrix element by matrix element,
\[
\bigl(e^{i\frac{F}{2}\mathbb{S}^3} P^+ e^{-i\frac{F}{2}\mathbb{S}^3}\bigr)_{mn}
= e^{i\frac{F}{2}(s_m-s_n)} (P^+)_{mn}.
\]
The right‑hand side gives $e^{i\frac{F}{2}}(P^+)_{mn}$.  
For every $(m,n)$ with $(P^+)_{mn}\neq0$ we cancel the non‑zero real factor $(P^+)_{mn}$ and obtain
\[
e^{i\frac{F}{2}(s_m-s_n)} = e^{i\frac{F}{2}}. \tag{A.2}
\]

For the $S^+$ part, $(m,n)=(m,m+1)$ with $s_m-s_{m+1}=1$.  
Then (A.2) becomes $e^{iF/2}=e^{iF/2}$, an identity.
For the $(S^-)^{L-1}$ part, the only non‑zero entry is $(m,n)=(L,1)$.  
Here $s_L-s_1 = -\frac{L-1}{2} - \frac{L-1}{2} = -(L-1)$.  
Equation (A.2) yields
\[
e^{-i\frac{F}{2}(L-1)} = e^{i\frac{F}{2}} \;\Longrightarrow\; e^{-i\frac{F}{2}L}=1.
\]
Hence
\[
\frac{F}{2}L = 2\pi n \;\Longrightarrow\; F = \frac{4\pi n}{L},\qquad n\in\mathbb{Z}. \tag{A.3}
\]

Thus periodic boundary conditions enforce the flux quantization $F=4\pi n/L$.  
For $L=3$ this gives $F=4\pi n/3$, for $L=4$ it gives $F=\pi n$, in agreement with the main text.  
The precise numerical values of the matrix elements of $\mathbb{S}^\pm$ (here chosen as $1$) cancel out in (A.2) and therefore do not affect the result.

\acknowledgments

This work was supported by the National Natural Science Foundation of China (Grant No. 12504192), the Shanghai Science and Technology Innovation Action Plan (Grant No. 24LZ1400800). YLW is supported by an appointment to the Young Scientist Training Programme at the APCTP through the Science and Technology Promotion Fund and Lottery Fund of the Korean Government.


\bibliographystyle{JHEP}
\bibliography{biblio.bib}

\end{document}